\documentclass[aps,twocolumn,nofootinbib,groupedaddress,amsfonts,floatfix]{revtex4-1} 
\pdfoutput=1
\usepackage{graphicx,amsmath,amssymb,amstext}
\usepackage{amssymb,amsbsy,amsfonts,amsthm,color}

\usepackage{epsfig}
\usepackage{graphicx}
\usepackage{subfigure}
\usepackage{hyperref}

\graphicspath{{Figures/}}

\usepackage{color}
\newcommand{\Beq}{\begin{eqnarray}}
\newcommand{\Eeq}{\end{eqnarray}}

\newcommand{\eqn}[1]{Eqn. (\ref{#1})}

\def\lsim{\mathrel {\vcenter {\baselineskip 0pt \kern 0pt \hbox{$<$} \kern 0pt \hbox{$\sim$} }}}

\newcommand{\mpl}{M_{\mbox{\tiny Pl}}}

\def\gsim{\mathrel {\vcenter {\baselineskip 0pt \kern 0pt \hbox{$>$} \kern 0pt \hbox{$\sim$} }}}

\newcommand{\grchombo}{\mathtt{GRChombo}}

\begin{document}
{\hfill UTTG-14-16, KCL-PH-TH/2016-53}
\vskip .5cm

\title{Robustness of Inflation to Inhomogeneous Initial Conditions}

\author{Katy Clough${}$}
\email{katy.clough@kcl.ac.uk}
\author{Eugene A. Lim${}$}
\email{eugene.a.lim@gmail.com}
\affiliation{${}$Theoretical Particle Physics and Cosmology Group, Physics Department,
Kings College London, Strand, London WC2R 2LS, United Kingdom}
\author{Brandon S. DiNunno${}$}
\email{bsd86@physics.utexas.edu}
\author{Willy Fischler${}$}
\email{fischler@physics.utexas.edu}
\author{Raphael Flauger${}$}
\email{flauger@physics.utexas.edu}
\author{Sonia Paban${}$}
\email{paban@physics.utexas.edu}
\affiliation{${}$ Department of Physics, The University of Texas at Austin, Austin, TX, 78712, USA}

\begin{abstract}
We consider the effects of inhomogeneous initial conditions in both the scalar field profile and the extrinsic curvature on different inflationary models. In particular, we compare the robustness of small field inflation to that of large field inflation, using numerical simulations with Einstein gravity in 3+1 dimensions.  We find that small field inflation can fail in the presence of subdominant gradient energies, suggesting that it is much less robust to inhomogeneities than large field inflation, which withstands dominant gradient energies. However, we also show that small field inflation can be successful even if some regions of spacetime start out in the region of the potential that does not support inflation. In the large field case, we confirm previous results that inflation is robust if the inflaton occupies the inflationary part of the potential. Furthermore, we show that increasing initial scalar gradients will not form sufficiently massive inflation-ending black holes if the initial hypersurface is approximately flat. Finally, we consider the large field case with a varying extrinsic curvature $K$, such that some regions are initially collapsing. We find that this may again lead to local black holes, but overall the spacetime remains inflationary if the spacetime is open, which confirms previous theoretical studies.

\end{abstract}

\maketitle
\section{Introduction}

Cosmic Inflation \cite{Guth:1980zm,Linde:1981mu,Albrecht:1982wi,Starobinsky:1980te} is thought to provide a solution to several problems in standard Big Bang theory by dynamically driving a ``generic'' initial state to a flat, homogeneous and isotropic Universe, while generating a nearly scale-invariant power spectrum of primordial perturbations which is consistent with observations. 
The question of what constitutes a ``generic'' initial state is a difficult one, and can only be understood in the context of a quantum theory of gravity. However, regardless of the nature of quantum gravity, a random realisation from the set of all possible initial conditions will not look like an inflationary spacetime, and one should expect the initial conditions from which inflation begins to contain some measure of inhomogeneity.

Inhomogeneities do not necessarily prevent inflation in models of chaotic inflation~\cite{Linde:1983gd} where inflation may naturally begin near the Planck scale~\cite{Linde:1984ir,Linde:1985ub,Linde:2014nna}. However, the simplest models of chaotic inflation in which inflation can start at the Planck scale are under pressure from recent observations of the cosmic microwave background~\cite{Array:2015xqh}. The data does not exclude scenarios in which inflation begins near the Planck scale~\cite{Mukhanov:2013tua,Kallosh:2014xwa}, but it motivates a study of the effects of inhomogeneities on scenarios in which the potential energy density is always sub-Planckian. While inflation in this class of models may start naturally if the topology of the spatial slices of our universe is non-trivial~\cite{Linde:2004nz}, or if the cores of topological defects serve as seeds~\cite{Linde:1994hy,Vilenkin:1994pv,Linde:1994wt}, the effects of inhomogeneities on the onset of inflation in these models in general is less understood and will be the focus of this work.

The issue of initial conditions for inflation and the stability of de Sitter and inflationary spacetimes have been under investigation for as long as inflation itself~\cite{Linde:1983gd,Linde:1985ub}, and there are many analytic and semi-analytic~\cite{Gibbons:1977mu,Hawking:1981fz,Wald:1983ky,Starobinsky:1982mr,Barrow:1984zz,Albrecht:1984qt,Barrow:1985,Gibbons:1986xk,Jensen:1986nf,Hawking:1987bi,Penrose:1988mg,Muller:1989rp,Kitada:1991ih,Kitada:1992uh,Bruni:1994cv,Maleknejad:2012as,Gibbons:2006pa,Boucher:2011zj,Bruni:2001pc,Muller:1987hp,Barrow:1989wp,Bicak:1997ne,Capozziello:1998dq,Vachaspati:1998dy,Barrow:1987ia,Barrow:1986yf,Polyakov:2009nq,Marolf:2010nz,Tsamis:1992sx,Brandenberger:2002sk,Geshnizjani:2003cn,Marozzi:2012tp,Brandenberger:1990wu,Carroll:2010aj,Corichi:2010zp,Schiffrin:2012zf,Remmen:2013eja,Corichi:2013kua,Mukhanov:2014uwa,Remmen:2014mia,Berezhiani:2015ola,Kleban:2016sqm} as well as numerical studies~\cite{Albrecht:1985yf,Albrecht:1986pi,KurkiSuonio:1987pq,Feldman:1989hh,Brandenberger:1988ir,Goldwirth:1989pr,Goldwirth:1989vz,Brandenberger:1990xu,Laguna:1991zs,Goldwirth:1991rj,KurkiSuonio:1993fg,Easther:2014zga,East:2015ggf,Braden:2016tjn,Alho:2011zz,  Alho:2013vva} (see~\cite{Brandenberger:2016uzh} for a short review). 
Goldwirth and Piran~\cite{Goldwirth:1989pr,Goldwirth:1989vz,Goldwirth:1991rj} were the first to study the robustness of inflation to spherically symmetric perturbations using general relativistic 1+1D simulations.\footnote{An earlier pioneering work \cite{Albrecht:1985yf} showed that inhomogeneous scalar fields will homogenise in a fixed FRW background. See also \cite{Easther:2014zga} for a recent follow up work in this direction.} In modern terminology, their conclusion was that large field models, in which the inflaton traverses more than a Planck mass during the inflationary period, $\delta \phi \gtrsim \mpl$, are more robust than small field models, $\delta \phi \ll \mpl$. Their results are often taken to imply that inflation requires a homogenous patch of size roughly~$1/H$ to begin. This work was later followed by 3+1D numerical simulations in Refs.~\cite{KurkiSuonio:1993fg,Laguna:1991zs} showing large field inflation to be robust to simple inhomogeneous (and anisotropic) initial conditions with large initial gradient energies in situations in which the field is initially confined to the part of the potential that supports inflation. This was confirmed recently in Ref.~\cite{East:2015ggf}, which demonstrated that large field inflation is robust even if $\rho_{\mathrm{grad}} \approx 1000\,\rho_{V}$ where $\rho_{V}$ is the vacuum energy density, at least if the universe initially expands at the same rate everywhere. Furthermore,~\cite{East:2015ggf}  first presented simulations in which inflation succeeded even for certain initial conditions that lead to the formation of black holes.

In this paper, we continue this line of research and test the robustness of inflation to a slightly more general but still very simple class of inhomogeneous initial conditions both in the scalar field profile and the extrinsic curvature. We use the numerical relativity package $\grchombo$~\cite{Clough:2015sqa}, setting up the machinery that will allow us to study more general classes of initial conditions in the future.  Since the degree of robustness to inhomogeneities depends on the exact model of inflation, this provides us with an approach to checking model viability. According to the Lyth bound~\cite{Lyth:1996im,Turner:1996ck}, inflation occurs at high energies and involves large field excursions in models that produce observable amounts of primordial gravitational waves, whereas the energy scale and field excursion are small in models that do not. 
Our results are summarized as follows:
\begin{itemize}
\item{For the initial conditions we consider, we find that large field inflation is robust to large gradient energies of $\rho_{\mathrm{grad}}/\rho_V \gg 1$, in agreement with~\cite{Laguna:1991zs,KurkiSuonio:1993fg,East:2015ggf}.}
\item{{Small field inflation is less robust than large field inflation.} It can fail even when the energy density in gradients is subdominant $\rho_{\mathrm{grad}}/\rho_V \ll 1$. We show that small field inflation fails when a large enough local fluctuation ends inflation early in that particular region, with the gradients quickly dragging the rest of the spacetime from the inflating part of the potential. However, the size of local fluctuation required to end inflation must be large enough to explore the boundary of inflationary regime of the potential, making small field inflation somewhat more robust than might be expected.}
\item{{Large inhomogeneities do not form dominant black hole spacetimes.} In the large field case, the potential is sufficiently wide to support inhomogeneities which result in collapse to form black holes. However, in the case where the initial spacetime is flat on average, increasing gradient energy implies an increase in average initial expansion. This expansion prevents the formation of inflation-ending black hole spacetimes. We found that there exists a maximum black hole mass which is subdominant to the inflationary spacetime, which we derived both analytically and numerically.  }
\item{We show that for initial spacetimes containing both expanding and collapsing regions local regions may collapse into black holes. However, inflation will occur as long as the spacetime is on average initially expanding. This is consistent with the theoretical expectations of \cite{Barrow:1985} and \cite{Kleban:2016sqm}.}
\end{itemize}

This paper is organised as follows. In Section \ref{sect:method} we present the theory and methodology of our approach. In Sections \ref{sect:SF} and \ref{sect:LF}, we present the numerical results, and discuss their implications for the small field and large field cases respectively. We conclude in Section \ref{sect:conclusions}.

Movies of several of the simulations described in this paper can be accessed via the $\grchombo$ website at \url{http://grchombo.org}.

\section{Theory and Methodology} \label{sect:method}

We consider single-field inflation with canonical kinetic term
\begin{equation}
L_{\phi} = -\frac{1}{2}g^{\mu\nu}\partial_\mu \phi\partial_\nu\phi - V(\phi) \label{eqn:singlefield},
\end{equation}
For a spatially homogeneous configuration, inflation occurs if $V>0$ and the slow-roll parameters satisfy 
\begin{equation}
\epsilon  = \frac{\mpl^2}{16\pi}\left(\frac{V'}{V}\right)^2 \ll 1~,~\eta = \frac{\mpl^2}{8\pi}\left|\frac{V''}{V}\right|^2 \ll 1\,,
\end{equation}
with  $\mpl^2 = \hbar c/G$.
In this case the field is slowly rolling and $V\approx \mathrm{constant}$ acts as a cosmological constant resulting in an inflating spacetime. The second condition $\eta \ll 1$ is required to ensure that inflation occurs for the sufficient amount of e-foldings. 

\begin{figure}
\begin{center}
\includegraphics[width=8cm]{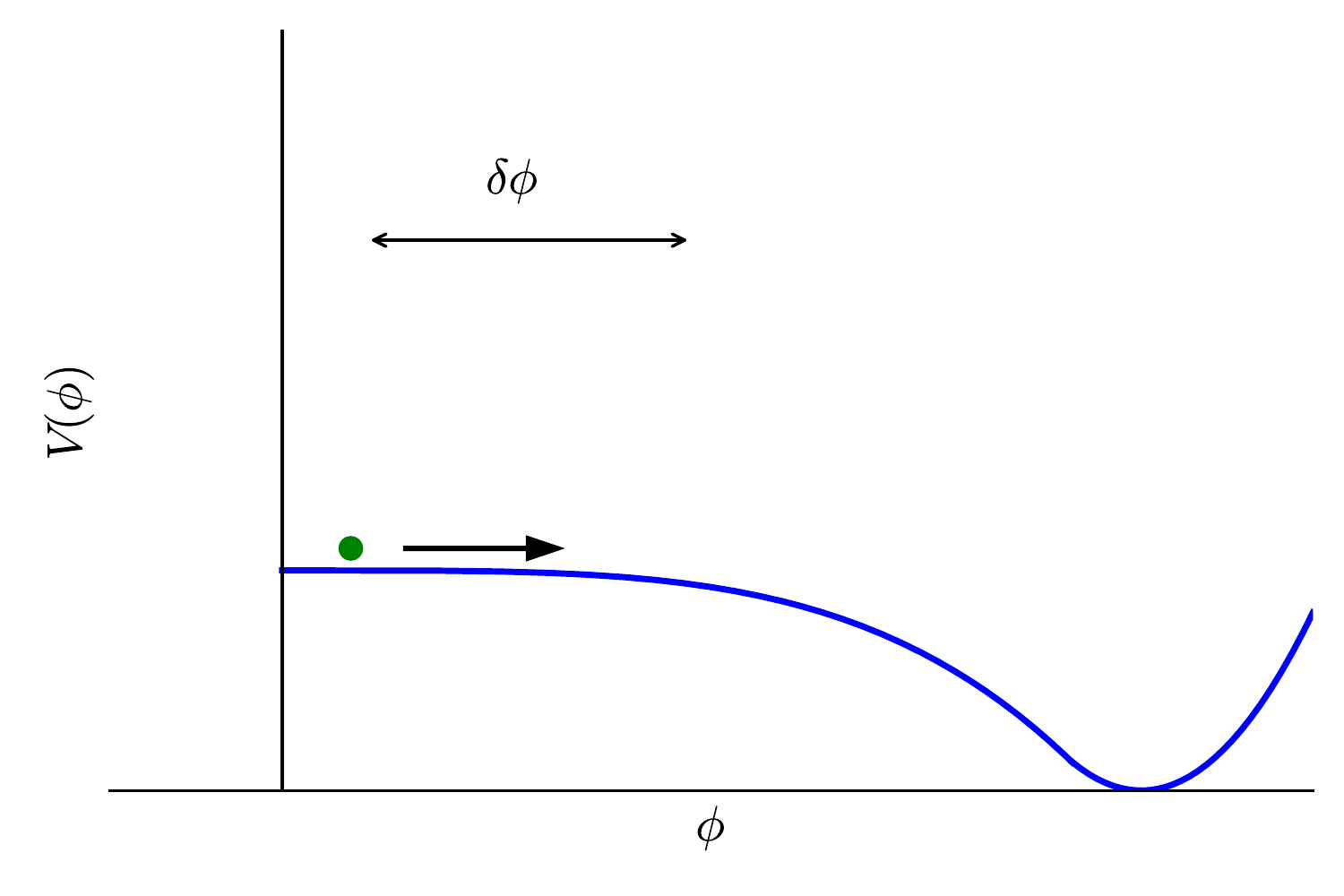}
\caption{In small field inflation the width of the inflationary ``slow-roll" part of the potential $\delta \phi \ll \mpl$, whereas in large field $\delta \phi \gg \mpl$.
\label{fig:SFvsLF}}
\end{center}
\end{figure}

In the large-field models, the region of the potential where this occurs is super-Planckian, i.e. the field needs to roll $\delta \phi \gtrsim \mpl$ for sufficient inflation, while in the small field model the field traverses a sub-Planckian distance in field space $\delta \phi  \ll \mpl$. This is illustrated in Figure \ref{fig:SFvsLF}. In the context of single field inflation, the Lyth bound~\cite{Lyth:1996im} implies that high/low-scale inflation is associated with large/small field inflation. 

\subsection{Initial Conditions}

\begin{figure}
\begin{center}
\includegraphics[width=8cm]{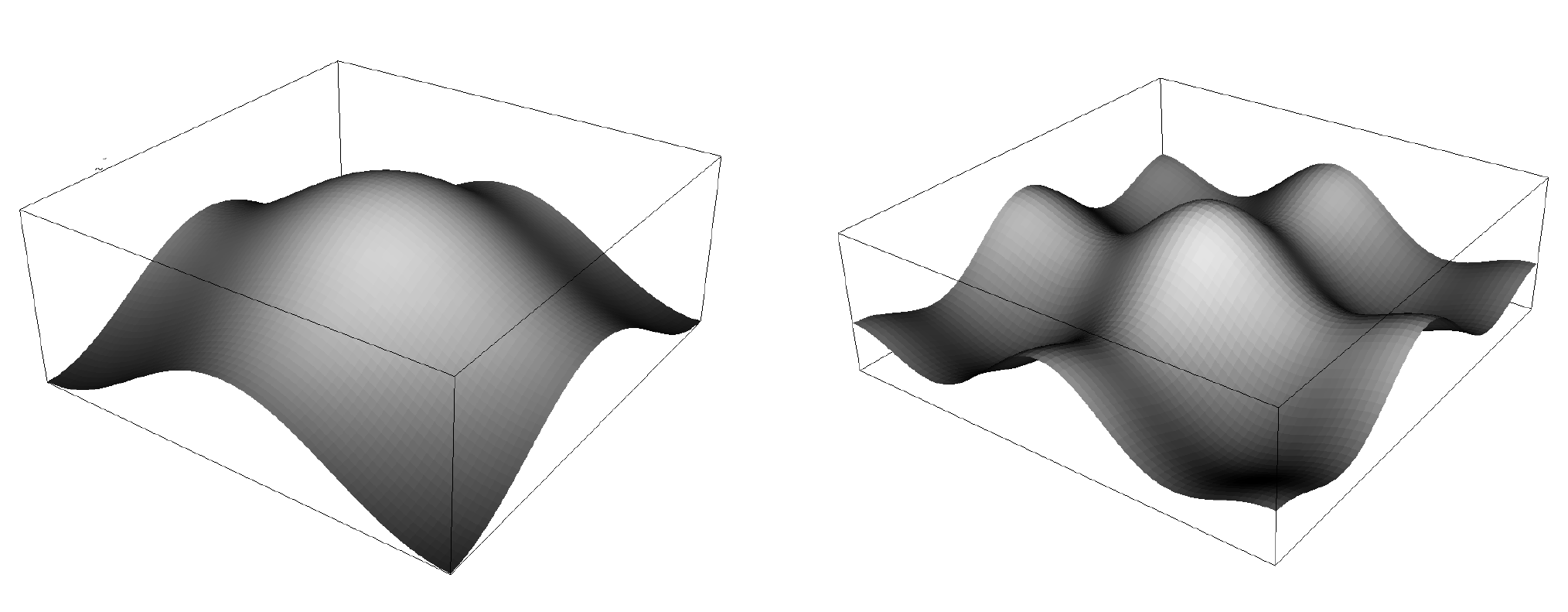}
\caption{Illustration of the cases $N=1$ and $N=2$ showing the values of $\phi$ on a 2D slice through the y axis. On this slice the maximum value of $\phi$ in each case is $4 \times 10^{-4} \mpl$ and the minima are $-9.5 \times 10^{-5} \mpl$ and $-6.0 \times 10^{-5} \mpl$ respectively.
\label{fig:Nis12}}
\end{center}
\end{figure}

We impose very simple inhomogeneous initial conditions similar to those in~\cite{East:2015ggf} by specifying the initial condition for the scalar field as follows
\begin{multline}
\phi(t = 0,{\bf x}) = \phi_0 \\ + \frac{\Delta \phi}{N} \sum_{n=1}^{N} \left( \cos{\frac{2 \pi nx}{L}} + \cos{\frac{2 \pi ny}{L}} + \cos{\frac{2\pi nz}{L}} \right) \label{eqn:phi}\,,
\end{multline}
and
\begin{eqnarray} \label{eqn:phidot}
\frac{\partial \phi(t = 0,{\bf x})}{\partial t} = 0 \,,
\end{eqnarray}
where ${\bf x}$ is the spatial coordinate of a foliation labeled by the time coordinate $t$, and $\Delta \phi$ is the amplitude of the initial inhomogeneities. The value $\phi_0$ is chosen such that we have 100 e-folds of inflation in the absence of any inhomogeneities. Since there are three modes each with amplitude $\Delta \phi$, the maximal total amplitude of the fluctuations about $\phi_0$ is $3\Delta \phi$. We chose not to include random phases in this work as we have found that random phases do not materially change the overall results. Note that we have normalised the total $\Delta \phi$ by the number of modes $N$ -- this means that the average gradient energy is slightly higher for larger $N$, but that the maximum traverse from $\phi_0$ towards the inflationary minimum is the same. See Figure \ref{fig:Nis12} for an illustration of the cases $N=1$ and $N=2$. 

We set $L$ to be the length of the simulation domain, and use periodic boundary conditions to simulate a space composed of periodic fluctuations of this length and amplitude. $L$ is chosen to be the Hubble length in the absence of inhomogeneities ($\Delta \phi=0$), that is
\begin{eqnarray}
L = \frac{3 \mpl}{\sqrt{24 \pi V(\phi_0)}} \label{eqn:Lformula}\,,
\end{eqnarray}
Hence, our model of initial inhomogeneities depends on the integer $N$, the amplitude of inhomogeneities $\Delta \phi$, and the potential $V(\phi_0)$. The potential $V(\phi_0)$ sets the inflationary Hubble scale, the integer $N$ sets the wavelength of the shortest perturbations relative to this scale, and $\Delta\phi$ sets the amplitude of the inhomogeneities. In this work we focus on $N$ of order unity and leave a more systematic study of the space of initial conditions for future work.
In large field inflation we use a single mode, i.e. $N=1$, and vary $\Delta \phi$. 
 In the limit in which the gradient energy dominates, i.e. $\rho_{\mathrm{grad}} \gg V(\phi_0)$, changing $\Delta \phi$ is equivalent to changing the wavelength of the mode relative to the actual Hubble length $H^{-1}$ (including the energy density from fluctuations). For $N=1$ and a Euclidean metric on the initial slice the wave number $k=2\pi/L$ satisfies 
\begin{equation}\label{eqn:equiv}
\frac{k}{H} = \frac{\mpl}{\sqrt{2\pi}\Delta \phi}\,.
\end{equation}
In small field inflation, we consider the two cases $N=1$, and $N=2$, a superposition of two modes, in addition to the variation of $\Delta \phi$. We will see that already small $\Delta \phi$ prevent successful inflation in the small field case so that we are not in the regime $\rho_{\mathrm{grad}} \gg V(\phi_0)$. 

We use the ADM metric, following the conventions of \cite{Clough:2015sqa},
\begin{equation}
ds^2=-\alpha^2\,dt^2+\gamma_{ij}(dx^i + \beta^i\,dt)(dx^j + \beta^j\,dt)\,,
\end{equation}
where $\alpha$ and $\beta^i$ are the lapse and shift parameters as usual. In the BSSN formalism of numerical relativity \cite{Baumgarte:1998te,Shibata:1995we}, these parameters are specified on the initial hypersurface and then allowed to evolve using gauge-driver equations. We choose $\alpha=1$ and $\beta_i=0$ on the initial hypersurface, and hence the initial gradient energy on this hypersurface is
\begin{equation}
\rho_{\mathrm{grad}} \equiv \frac{1}{2}\gamma^{ij}\partial_i \phi \partial_j \phi \label{eqn:grad}.
\end{equation}
It is convenient to introduce the conformal metric \mbox{$\tilde\gamma_{ij} =\chi \gamma_{ij}$} such that $\det \tilde{\gamma}_{ij} =1$, the corresponding Christoffel-symbols $\tilde\Gamma^i_{jk}$, as well as the short-hand notation $\tilde\Gamma^i=\tilde\gamma^{jk}\tilde\Gamma^i_{jk}$.
We can then write the equations we use to evolve the lapse and shift in the moving puncture gauge \cite{vanMeter:2006vi,Campanelli:2005dd}, which allows us to stably form and evolve black holes in the spacetime, as 
\begin{eqnarray}
\partial_t \alpha &=& -\mu_{\alpha}\alpha K + \beta^i\partial_i \alpha \ , \label{eqn:alphadriver}\\
\partial_t \beta^i &=& B^i\, \label{eqn:betadriver},\\
\partial_t B^i &=& \frac{3}{4} \partial_t \tilde\Gamma^i-\eta_B B^i\ \,. \label{eqn:gammadriver}
\end{eqnarray}
The exact values of $\mu_{\alpha}$ and $\eta_B$ are chosen to improve stability in any particular numerical simulation.

Next, we have to specify the initial conditions for the metric $\gamma_{ij}$ and extrinsic curvature $K_{ij}$. Their values need to satisfy both the Hamiltonian and momentum constraints on the initial hypersurface.
Decomposing the extrinsic curvature as
\begin{equation}
K_{ij} = \chi^{-1}\tilde{A}_{ij} + \frac{1}{3}\gamma_{ij}K~,~K = \gamma^{ij} K_{ij},
\end{equation}
and introducing the notation for the kinetic term
\begin{equation}
\eta \equiv \frac{1}{\alpha}\left(\partial_t{\phi} - \beta^k\partial_k \phi\right),
\end{equation}
so that the energy density at any point in the hypersurface is 
\begin{equation}
\rho = \frac{1}{2}\eta^2 + \frac{1}{2}\gamma^{ij}\partial_i \phi \partial_j \phi + V\,,
\end{equation}
the constraint equations become
\begin{multline}
\tilde{D}^2\chi -\frac{5}{4 \chi}\tilde{\gamma}^{ij}\tilde{D}_i\chi \tilde{D}_j\chi \\ + \frac{\chi \tilde{R}}{2} + \frac{K^2}{3} - \frac{1}{2} \tilde{A}_{ij}\tilde{A}^{ij} = 8\pi G \rho \label{eqn:HamCon}\,,
\end{multline}
and
\begin{equation}
\tilde{D}_j \tilde{A}^{ij} -\frac{3}{2 \chi} \tilde{A}^{ij}\tilde{D}_j \chi -\frac{2}{3}\tilde{\gamma}^{ij}\tilde{D}_j K = 8\pi G \eta \tilde{\gamma}^{ij} \partial_j \phi\,. \label{eqn:MomCon}
\end{equation}
$K$ is the local expansion rate of spacetime, and in the special case of the Friedmann-Robertson-Walker metric, $K = -3H$ where $H$ is the Hubble constant.\footnote{Note that we have chosen convention such that $K < 0$ denotes positive expansion.}

This is a set of coupled elliptic equations and is non-trivial to solve in general. Throughout this work, we will make the simplifying assumption that the metric is conformally flat and the traceless part of the extrinsic curvature $K_{ij}$ is zero everywhere on the initial slice
\begin{equation}
\tilde{\gamma}_{ij} = \delta_{ij}\,,  \label{eqn:conformalcond}
\end{equation}
and
\begin{equation}
\tilde{A}_{ij} = 0\,. \label{eqn:tracelessK}
\end{equation}
In this special class of initial conditions, we consider two possible solutions, that of uniform initial expansion $K$, and one with spatially varying $K$.

\subsubsection{$K=\mathrm{constant}$ uniform initial expansion}

\noindent For spatially varying $\phi$,  the momentum constraint  \eqn{eqn:MomCon} is trivially satisfied for $\eta=0$ and $K=$ const. $K$ is in principle a free parameter, corresponding to a uniform local expansion rate across the initial hypersurface. However, in order to satisfy periodic boundary conditions for $\chi$ and the Hamiltonian constraint, $K^2/24 \pi$ needs to lie close to the average initial energy density for the hypersurface. For simplicity, we choose it to be equal to the average initial energy density, approximating the metric to be Euclidean 
\begin{equation}
K = -\sqrt{24 \pi G \langle \rho \rangle}\,, \label{eqn:Kavg}
\end{equation}
with
\begin{equation}
\rho = \frac{1}{2}(\partial_i \phi)^2 + V(\phi)\,,
\end{equation}
where $\langle X \rangle = {\cal V}^{-1} \int X~d{\cal V}$ indicates the average over the spatial volume ${\cal V}$ of the quantity $X$.  Once $K$ is chosen, the initial field profile and the Hamiltonian constraint then fully determine the conformal factor $\chi$ (which we solve for using numerical relaxation).

In cases where the gradient energy dominates i.e. $\rho \approx \rho_{\mathrm{grad}} \gg V(\phi)$, the initial expansion rate is large compared to the Hubble rate associated with inflation. This large initial uniform expansion means that we are stacking the deck against ending inflation. In general, we should expect the local expansion rate to be a function of spatial position that can be both initially expanding or collapsing. To study the general case will require relaxing some combination of the conformal condition \eqn{eqn:conformalcond}, the condition on $A_{ij}$, \eqn{eqn:tracelessK}, and the condition of zero initial scalar field velocity, $\eta = 0$. We reserve the general case for future work, but there exists a second solution consistent with equations~\eqref{eqn:conformalcond} and~\eqref{eqn:tracelessK}, given our assumptions, from imposing $\eta \neq 0$, which gives non-uniform initial expansion. We turn to this solution next.

\subsubsection{$K\neq \mathrm{constant}$ expanding/contracting initial condition}

\noindent For constant initial scalar velocity $\eta$
\begin{equation}
\eta = -\frac{C}{12\pi G}, \label{eqn:etaansatz}
\end{equation}
with $C$ some constant, the momentum constraint~\eqn{eqn:MomCon} relates the extrinsic curvature $K$ to the initial scalar field profile $\phi$
\begin{equation}
K = -C\phi + K_0, \label{eqn:Kansatz}
\end{equation}
where $K_0$ is an integration constant.  This initial condition means that a \emph{constant} initial scalar velocity $\eta$ and a varying field $\phi$ will lead to a spatially varying $K$. If $K_0$ is chosen to be approximately the average value of $C \phi$, the spacetime will be locally initially expanding or contracting depending on its position. We can then again solve the Hamiltonian constraint for the conformal factor $\chi$ in order to complete  the specification of the initial conditions. 

\subsection{Numerical Set-up}

We rescale our simulations (by choosing the geometrized mass unit $M$ to represent some convenient fraction of $\mpl$) such that the size of our physical domain is covered by $(32M)^3$. We turn on $\grchombo$'s adaptive mesh refinement, using the gradients of $K$ and $\phi$ as refinement threshold conditions, with a coarsest level grid size of $64^3$, allowing up to 6 of levels of refinement with a refinement ratio of $2$ per level. We check convergence approximately in this case by checking that the same results are obtained when starting from a coarsest grid of $128^3$, increasing the number of grids by one and using a more aggressive regridding condition (approximately halving the thresholds). It was found that the difference in the results was small -- for example, the number of $e$-folds at failure in the small field cases were different by $\pm 0.1\%$.

We can track inflationary simulations for around 23 \mbox{$e$-folds}. After this point numerical error begins to dominate as the conformal factor $\chi$ (equal to the inverse of the square of the scale factor) falls below working precision.

\section{Small Field Inflation} \label{sect:SF}

\begin{figure}
\begin{center}
\includegraphics[width=8cm]{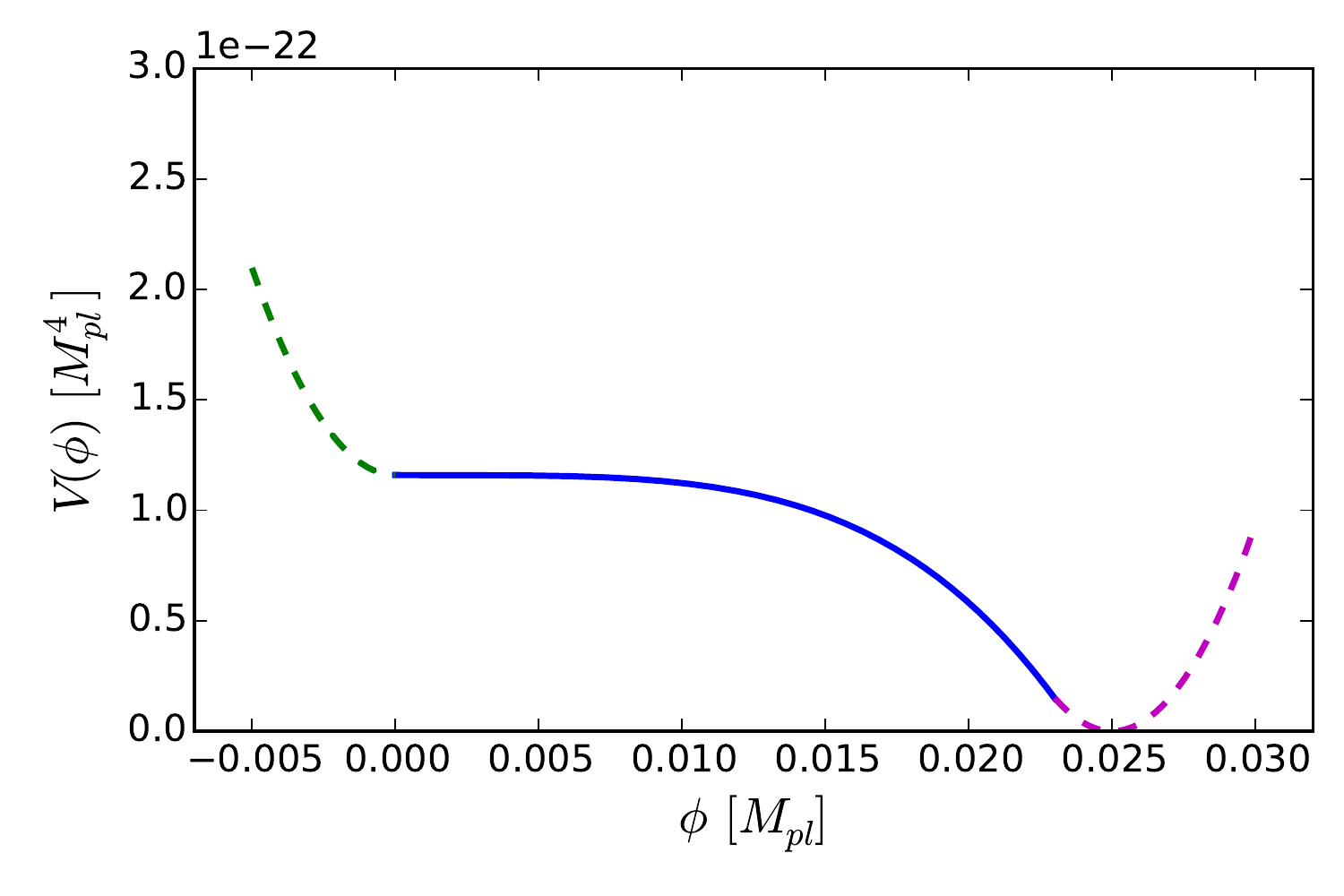}
\caption{Small field potential function $V(\phi)$ without an extended flat region, showing the three regions, the central solid line (blue) region gives rise to the slow-roll inflationary period
\label{fig:PotSFsteep}}
\end{center}
\end{figure}

As discussed in the Introduction, the inflating plateau of the small field potential can be relatively narrow, with $\delta \phi \ll \mpl$. The reason is as follows. The scalar power spectrum for single field inflation is given by 
\begin{equation}
\Delta_R^2=\frac{H^2}{\pi \mpl^2 \epsilon} \approx 2\times 10^{-9} \label{eqn:Ps}\,.
\end{equation}
For low scale inflation $(H/\mpl)^2$ is small, which means that $\epsilon$ must be small to achieve sufficient amplification of the observed scalar power. This means that the inflaton has to be roll very slowly (when compared to the large field case). The number of e-folds $\mathcal{N}$, is given by 
\begin{equation}
\mathcal{N} \approx  \int \sqrt{\frac{4\pi}{\epsilon}} \frac{|d\phi|}{\mpl}, \label{eqn:Nfolds}\,.
\end{equation}
Assuming $\epsilon$ and $H$ constant, we estimate the field range with the help of equation~\eqref{eqn:Ps} 
\begin{equation}
\delta \phi \approx \frac{\mathcal{N}}{2}\frac{H}{\mpl}10^{5} \mpl. \label{eqn:totaltraverse}
\end{equation}
For a typical low scale inflation $H/\mpl \sim 10^{-10}$, we see that $\delta \phi/\mpl\sim 3 \times 10^{-4}$ so $\delta \phi$ is sub-Planckian as we argued. A typical small field inflation model is shown in Fig. \ref{fig:PotSFsteep}, where the inflating domain is around the inflection point of the potential.

\begin{figure}
\begin{center}
\includegraphics[width=8cm]{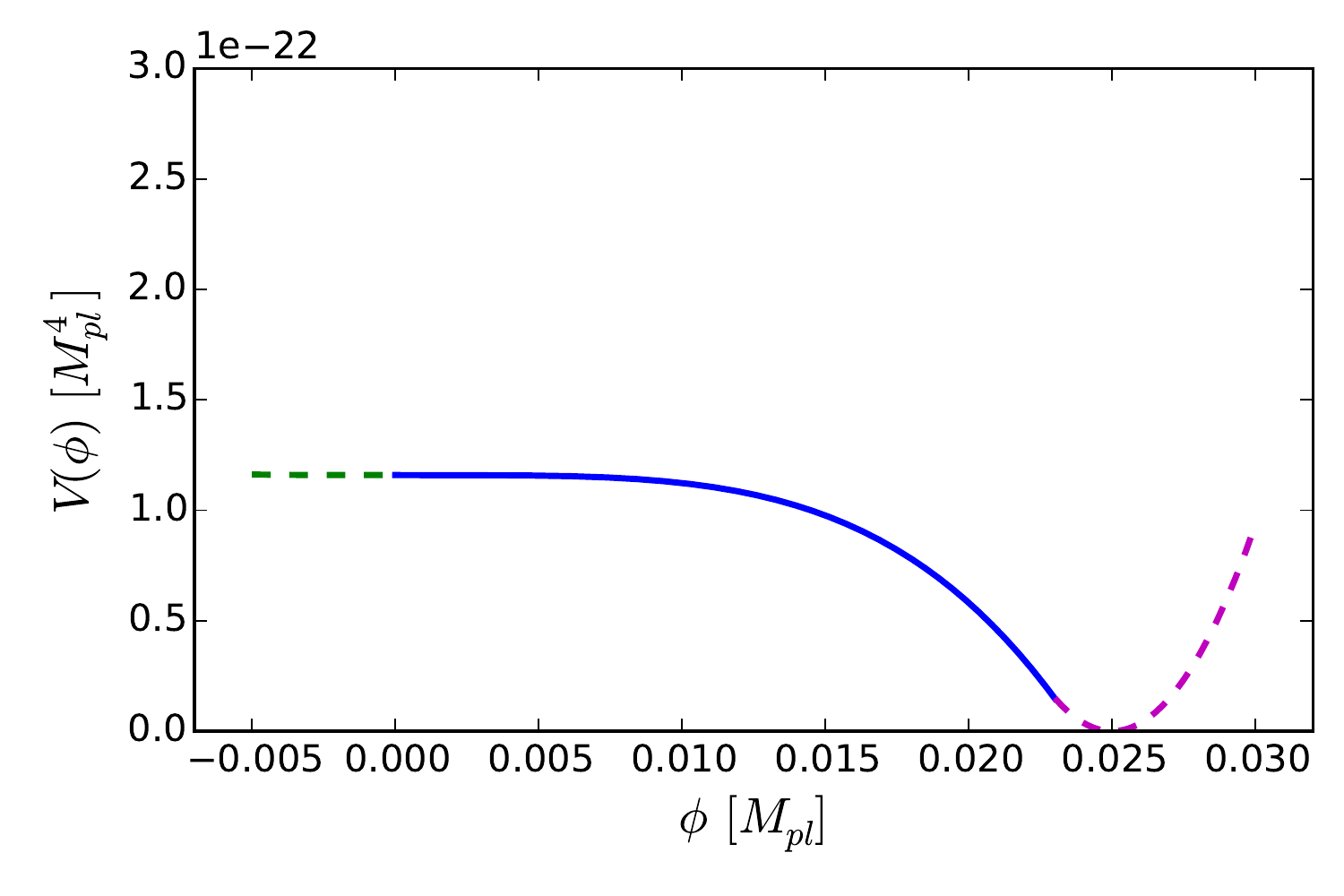}
\caption{Small field potential function $V(\phi)$ with an extended flat region, showing the three regions, the central solid line (blue) region gives rise to the slow-roll inflationary period
\label{fig:PotSFflat}}
\end{center}
\end{figure}

Inflation could occur for much longer than 60 \mbox{$e$-folds}. Therefore the inflection point could be quite broad, while still accommodating the small field requirement, so the potential could instead look like Figure~\ref{fig:PotSFflat}, with a wider plateau.  In the context of inhomogeneous inflation, this distinction is important -- the presence of large gradients means that the scalar field can now sample a large domain of the potential, which could include the non-inflating ``cliff'' on the left side of the potential in Figure~\ref{fig:PotSFsteep}. The additional potential energy in such regions is converted to scalar kinetic energy as the field rolls down the hill towards the inflection point, which can disrupt slow roll sufficiently to end inflation.

In this section, we will explore and compare the two cases, a potential with an extended flat direction, and one with a steeper rise. Note that we do not consider the effect of varying $K$ on small field inflation. This is because a profile for $K$ which covers both negative and positive values (i.e. with both expanding and contracting regions), requires the addition of a relatively large kinetic energy term $\eta$, which immediately pushes the field into the minimum, ending inflation. Thus the case where $K$ is constant represents a best case scenario -- adding variation in $K$ will only end inflation sooner. This is consistent with initial kinetic energy being the most important failure mode, as shown by Goldwirth and Piran.

\subsection{Small Field model with extended flat direction} \label{sec-smallflat}

In this section, we will investigate the robustness of small field inflation for the case depicted in Figure \ref{fig:PotSFflat}.

We model the inflationary potential as
\begin{equation}
V(\phi)=
\begin{cases}
V_0& \phi<0\\
V_0 \left(1 - \left(\frac{\phi}{\mu}\right)^4\right) & 0<\phi<0.023 \mpl\\
m^2 (\phi - \phi_*)^2 & \phi > 0.023 \mpl
\end{cases}
\end{equation}
with $\mu = 0.0238\mpl$, $V_0 = 1.15949 \times 10^{-22} \mpl^{4}$, \mbox{$m^2 = 3.75 \times 10^{-18} \mpl^2$} and $\phi_* = 0.025\mpl$. The Hubble rate during inflation for this choice of parameters is $H_{\rm inf} = 3.125 \times 10^{-11} \mpl$.

For a homogeneous initial value of the field of $\phi_0 = 10^{-4} \mpl$, these values would result in 100 e-folds of inflation, $\Delta_R = 10^{-5}$ and $n_s = 0.95$ for modes that exit the horizon 60 $e$-folds before the end of inflation. We find the end of the inflationary plateau, the point at which the potential is no longer ``slow roll", to be at approximately $\phi=0.008 \mpl$, with all but the last $e$-fold taking place for $\phi<0.001\mpl$.  

The length scale for the fluctuations $L$, set to the Hubble length in the absence of fluctuations, is then \mbox{$L=3.2 \times 10^{10} \mpl^{-1}$}, and the value of $K$ is constant across the grid as described in section \ref{sect:method}. This satisfies the Hamiltonian constraint, assuming that the initial value of the conformal factor of the metric, $\chi$, is approximately of order 1. In our simulations, we set this constant value of $K$ across the grid and then relax the value of $\chi$ from a value of 1 everywhere to satisfy the Hamiltonian constraint exactly.\footnote{Although in the small field case $\chi$ remains very close to 1 as the fluctuations are small, and the space is approximately flat.}

\begin{figure}
\begin{center}
\includegraphics[width=8cm]{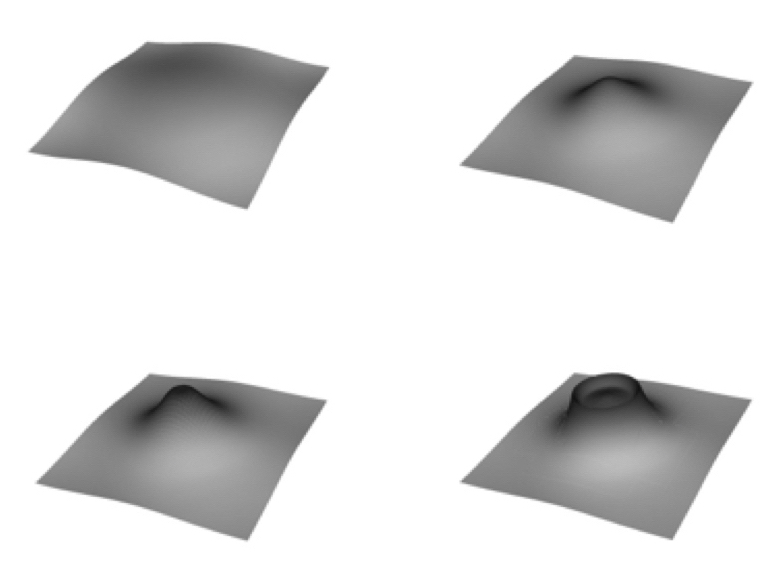}
\caption{A time series of the value of the field $\phi$ is shown on a 2 dimensional spatial slice. The maximum initial value of the field is at the centre, and it can be seen how this point ``falls off'' the inflationary potential, and subsequently drags the remaining space down with it. This means that the failure of a single point quickly ends inflation throughout the spacetime.
\label{fig:FallSpread}}
\end{center}
\end{figure}

\begin{figure}
\begin{center}
\includegraphics[width=8cm]{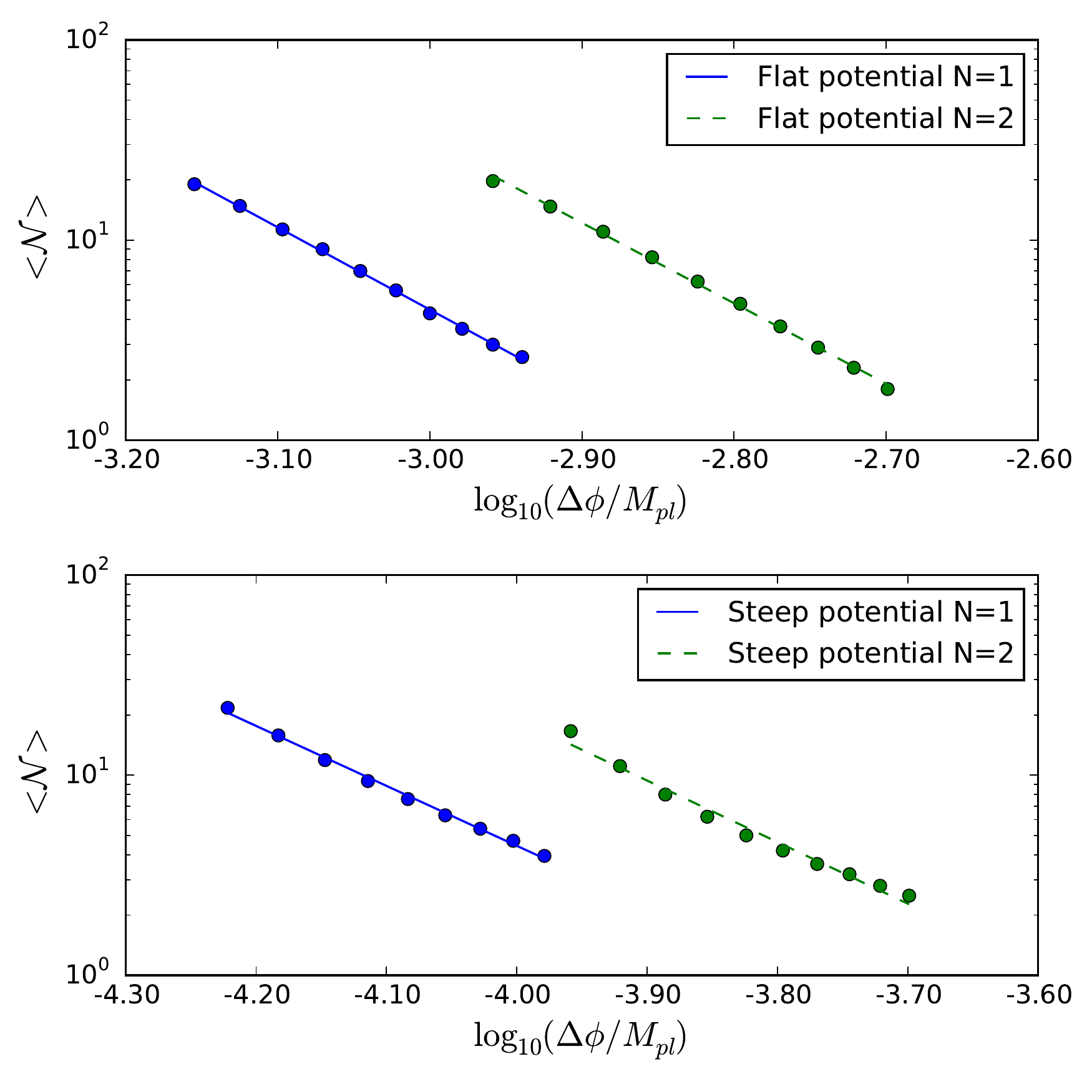}
\caption{Plot showing the failure of inflation for small field inflation in the steep and flat cases for $N=1$ and $N=2$. The average number of e-folds $\langle \mathcal{N} \rangle$ decreases as $\Delta \phi$ is increased, with an approximate relationship of $\langle \mathcal{N} \rangle \propto \Delta \phi^{-4}$ for the flat case and of $\langle \mathcal{N} \rangle \propto \Delta \phi^{-3}$ for the steep case.
\label{fig:NefoldsSF}}
\end{center}
\end{figure}

We then evolve the initial conditions forward in time until inflation ends, or we reach the maximum number of $e$-folds we can simulate. We define the end of inflation as being the point at which a single point in the space falls to the minimum of the potential, that is, when the value of $\phi = \phi_*$ somewhere on the grid. The rest of the space will subsequently be pulled in by gradients, as illustrated in Figure \ref{fig:FallSpread}, and as we will discuss in more detail in the next section. The average number of $e$-folds $\langle \mathcal{N} \rangle$ is measured on this time slice.\footnote{While the remaining spacetime can achieve several more e-folds before falling to the minimum, we treat this point as having ended inflation for measurement purposes. Allowing the simulations to run until the whole spacetime has fallen to the minimum and fully ceased inflating would displace the lines in Figure \ref{fig:NefoldsSF} vertically, but the trends would be the same. The actual values of $\langle \mathcal{N} \rangle$ are, in any case, model specific.} We do this for a range of $\Delta \phi$. The results are shown in Figure \ref{fig:NefoldsSF} for the cases $N=1$ and $N=2$. 

For $N=1$ we find that inflation ends with less than 20 e-folds (which we call ``failure" for our purposes) for initial amplitudes of around $\Delta \phi> 0.0007\mpl$. These values of $\Delta \phi$ correspond to $\rho_{\mathrm{grad}}/\rho_{V_0} \geq 1\times 10^{-4} $.\footnote{This implies that we are in a very different regime in our simulations of small-field models from~\cite{East:2015ggf} where the gradient energy density dominated the potential energy density in both the large- and small-field case. We find that already subdominant energy density in gradients can significantly impede or prevent inflation. } Hence the gradient energy is still sub-dominant to the length scale $L \approx 1/H$. This is highly ``homogenous''. Note that the density contrast in inflationary primordial perturbations are expected to be of order $10^{-5}$ which is only an order of magnitude smaller than this. However, the perturbations here are concentrated in one or two modes and it is the field excursions that should be compared. The typical displacement due to quantum fluctuations is $\Delta\phi_{\rm QM}\sim H_{\rm inf}/2\pi\ll0.0007\mpl$. We will explore the robustness of inflation to initial perturbations with more general power spectra in future work.

For $N=2$ we find failure when $\Delta \phi > 0.0011\mpl$. It can be seen that adding the additional mode makes inflation more robust. Recall from the definition \eqn{eqn:phi} that we have normalised $\Delta \phi$ to the total number of modes so adding modes adds to the gradient energy but not to the maximum field value -- \emph{this suggests that inflationary failure scenarios are more dependent on single long wavelength inhomogeneities rather than multiple short wavelength ones}. The number of $e$-folds decreases with an approximate relationship of $\langle \mathcal{N} \rangle \propto \Delta \phi^{-4}$ in both the $N=1$ and $N=2$ cases. 

\subsection{Pull back effects in small field inflation}

\begin{figure}
\begin{center}
\includegraphics[width=8cm]{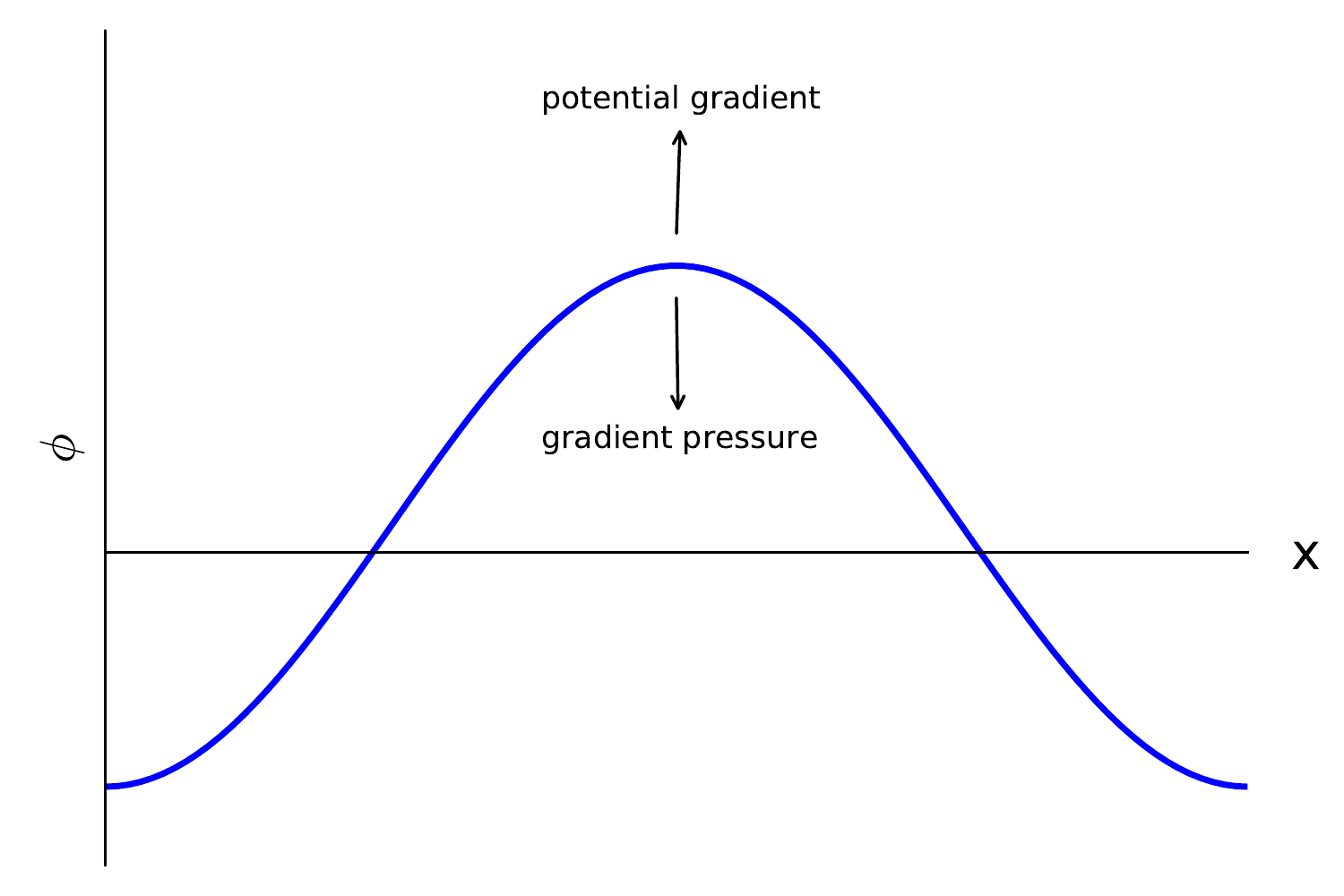}
\caption{Illustration of competition between gradient pressure and potential gradient for a concave model.}
\label{fig:Compete}
\end{center}
\end{figure}

As was mentioned above, once one part of the field falls into the minimum, it quickly ``drags down" the remaining spacetime, as shown in Figure \ref{fig:FallSpread}. It is instructive to consider the scalar field dynamics which leads to the failure as the naive expectation that the part of the field which has the maximum initial value (and hence is closer to the point where inflation ends) is that which falls to the minimum first is not always correct. There is some initial resistance from gradient pressure which, for a range of $\Delta \phi$, pulls the field back up the hill away from the minimum, ``saving" inflation and making it more robust to inflation than one might expect. See Figure \ref{fig:Compete}.

Using the Klein-Gordon equation, we can show that local gradient pressure should temporarily ``save" inflation up to some critical value, above which the field will fall directly to the minimum. The critical value for $\Delta \phi$ where this happens can be approximated quite accurately as follows. Consider the Klein-Gordon equation,
\begin{eqnarray}
\partial_t^2{\phi} - \gamma^{ij}\partial_i \partial_j\phi + \frac{d V}{d \phi}=0,
\end{eqnarray}
where we have ignored the friction term, and let 
\begin{equation}
\phi_{\max}(t) = \mathrm{max}(0,\phi({\bf x},t)).
\end{equation}
Initially, $\phi_{\rm max} = \phi_0 + 3 \Delta \phi$ and $\gamma^{ij}\sim {\cal O}(1)$. For this point, and assuming that we are still in the concave part of the potential, the field value should fall initially towards the minimum if 
\begin{eqnarray}
-\nabla^2 \phi  \lsim \left|\frac{dV}{d\phi}\right|.
\end{eqnarray}
For the initial conditions and the potential we consider, this is the case (assuming $N=1$) for
\begin{eqnarray}
3 k^2 \Delta \phi \lsim -\frac{dV}{d\phi} \bigg|_{\phi=\phi_{\rm max}}. \label{eqn:pullback}
\end{eqnarray}
Using the relation \eqn{eqn:Lformula}, $\phi_0 \ll 3 \Delta \phi$, and $V(\phi_0) \approx V_0$ for our small field case this becomes
\begin{eqnarray}
\frac{32 \pi^3 n^2 V_0 \Delta \phi}{\mpl^2} \lesssim \frac{4 V_0 (3 \Delta \phi)^3}{\mu^4},
\end{eqnarray}
which simplifies to
\begin{eqnarray}
\Delta \phi \gtrsim \sqrt{\frac{8 \pi^3}{27}} \frac{n \mu^2}{\mpl}.
\end{eqnarray}
The critical value for our chosen values and $n=1$ is $\Delta \phi \approx 0.0017\mpl$, which corresponds to $\phi_{\rm max, crit} =0.00525\mpl$, beyond the part of the potential that supports an extended period of inflation. Small field inflation is thus more robust than on might naively expect because local excursions towards the edge of the inflationary plateau are pulled back onto it. 

\begin{figure}
\begin{center}
\includegraphics[width=8cm]{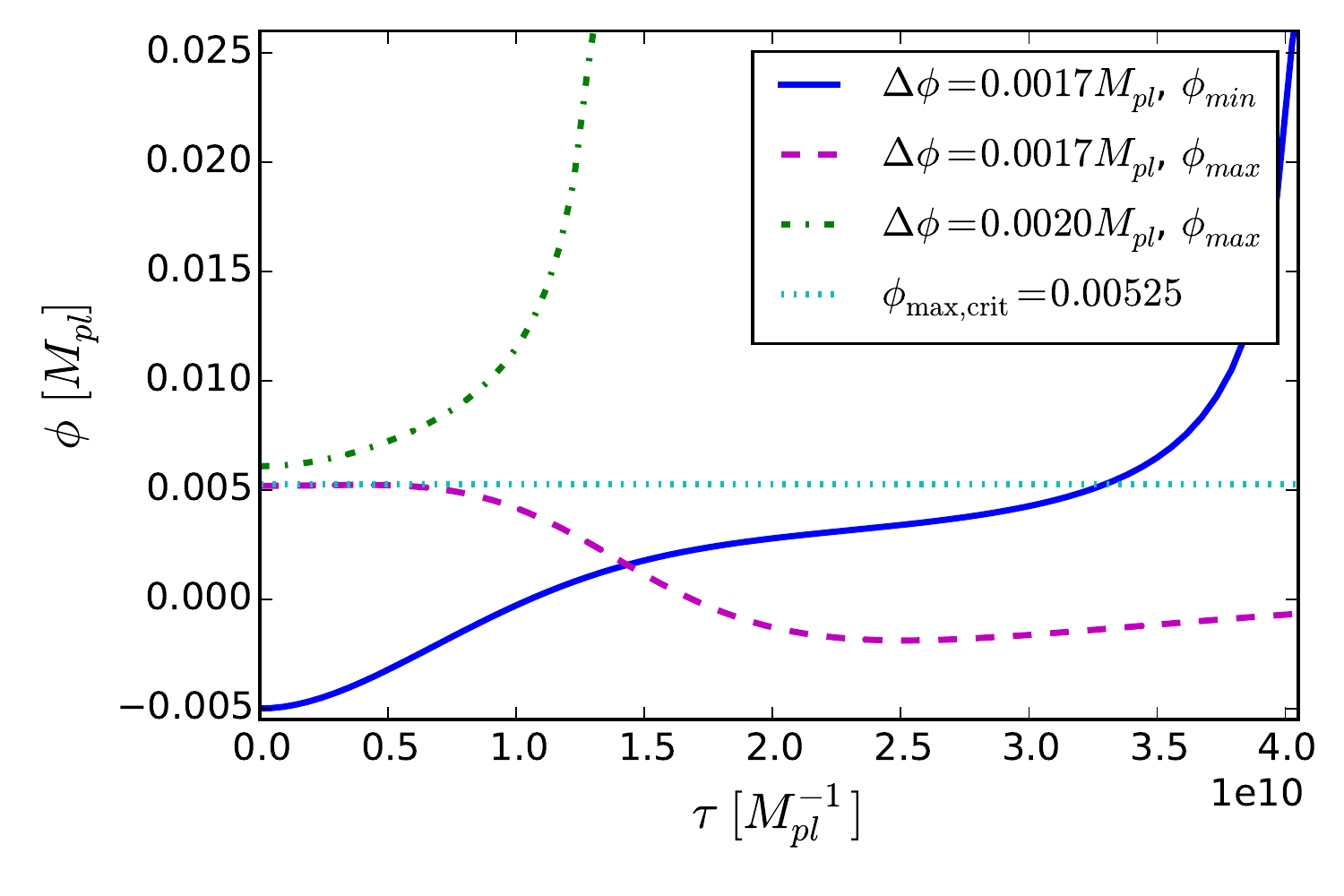}
\caption{The evolution of \emph{initial} $\phi_{\max}$ and $\phi_{\min} = \min(\phi({\bf x},t_0),+\infty)$  points versus proper time $\tau$ for small field inflation (extended flat case). At the derived ``critical value" of $\Delta \phi =0.0017\mpl$ and $\phi_{\rm max, crit} = 0.00525\mpl$, shows a very small initial increase at the maximum value (in dashed/pink) but then is pulled back, so that it is in fact the initial minimum point (shown in blue/solid) which fails first. A slightly larger value of $\Delta \phi =0.002\mpl$ fails immediately at the maximum point (in green/dotted), as expected. Note that $\tau$ is the proper time experienced by an observer at the specified coordinate location. 
\label{fig:SFphi_vs_t}}
\end{center}
\end{figure}

The results of several simulations are illustrated in \mbox{Figure~\ref{fig:SFphi_vs_t}}. We see that the predicted critical value for $\Delta \phi$ at which immediate failure occurs is approximately correct. In fact the field can even resist some initial movement towards the minimum at the maximum point, before being pulled back up the potential hill. 

\subsection{Small Field model without extended flat direction} \label{sec-smallsteep}

In this section, we will investigate the robustness of small field inflation in the case in which the negative $\phi$ direction is a ``cliff''. 
We model the potential as
\begin{equation}
V(\phi)=
\begin{cases}
V_0+m^2\phi^2& \phi<0\\
V_0 \left(1 - \left(\frac{\phi}{\mu}\right)^4\right) & 0<\phi<0.023 \mpl\\
m^2 (\phi - \phi_*)^2 & \phi > 0.023 \mpl
\end{cases}
\end{equation}
with the same parameters and initial conditions as in Section \ref{sec-smallflat}. 
The is illustrated in Figure \ref{fig:PotSFsteep}.

For the $N=1$ case, we find failure for initial amplitudes of $\Delta \phi \gtrsim 5 \times 10^{-5}\mpl$, which corresponds to $\rho_{\mathrm{grad}}/\rho_{V_0} \approx 6 \times 10^{-7} $. Note that inflation now fails for amplitudes roughly an order of magnitude smaller than in the case of the extended flat region, showing that small field inflation is highly sensitive to changes in the potential around the flat region. Although the energy density in these fluctuations is now smaller than those expected from inflationary primordial perturbations, recall that it is concentrated in one mode, rather than being a scale invariant spectrum. In particular, the typical displacement due to quantum fluctuations for a given mode is still significantly smaller than the fluctuations considered here $\Delta\phi_{\rm QM}\sim H_{\rm inf}/2\pi\ll 5\times 10^{-5}\mpl$. 

The results are shown in Figure \ref{fig:NefoldsSF} for the cases $N=1$ and $N=2$, below those for the case with the flatter potential. Again it can be seen that adding the additional mode makes inflation slightly more robust. The number of $e$-folds decreases as a power law with an approximate relationship of $\langle \mathcal{N} \rangle \propto \Delta \phi^{-3}$.

\begin{figure}
\begin{center}
\includegraphics[width=8cm]{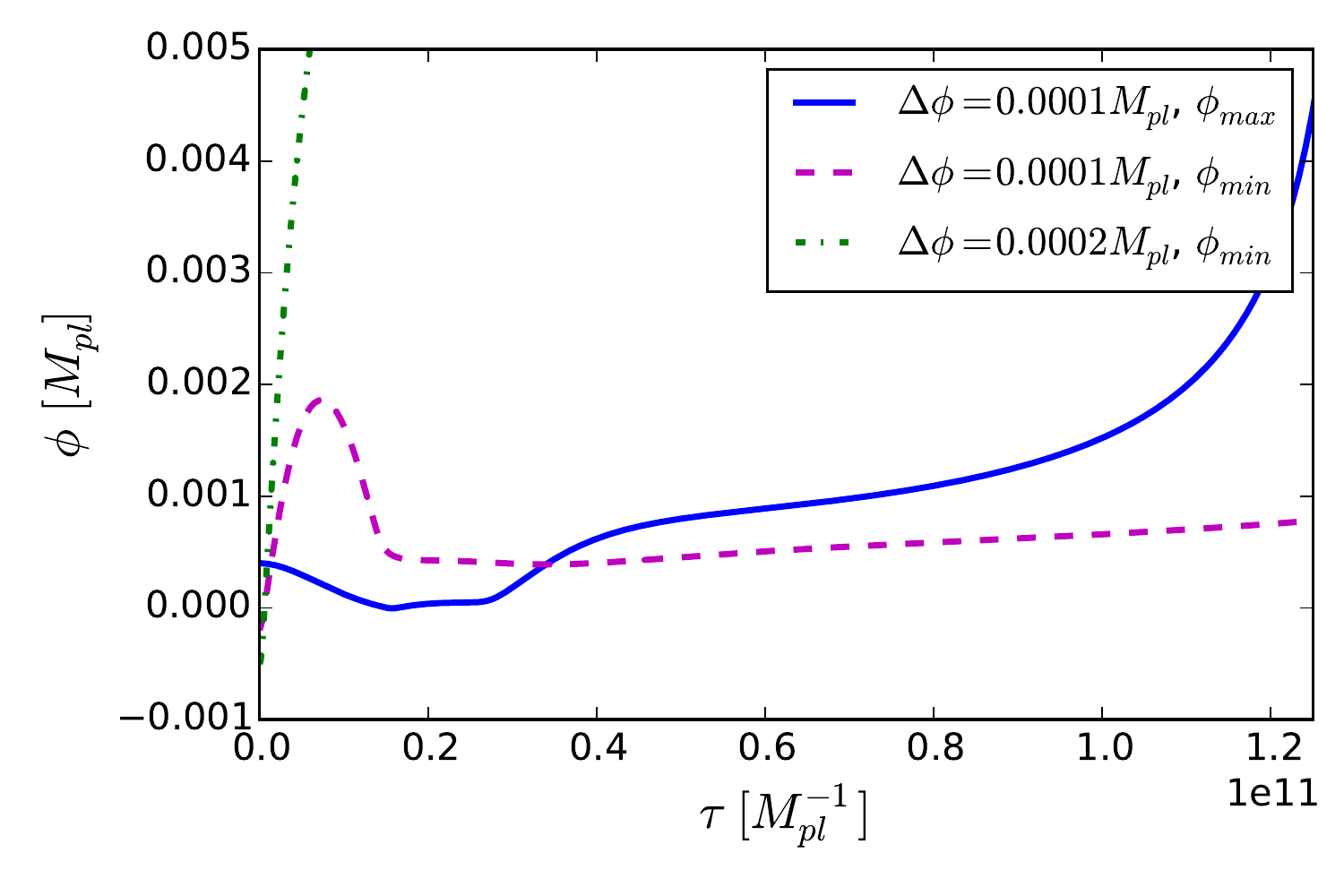}
\caption{The evolution of $\phi$ vs proper time $\tau$ for the initial minimum and maximum points, for the steep potential. The initial maximum point fails first, after hitting the ``wall'' to the left of the inflationary potential, and rebounding into the minimum. While the most negative point does rapidly overshoot due to its kinetic energy, it is pulled back by gradients and ultimately ``saved" from failure. However, increasing $\Delta \phi$ further does eventually lead to the most negative point dominating the collapse due to its kinetic energy. Note that $\tau$ is the proper time experienced by an observer at the specified coordinate location. 
\label{fig:SFphi_vs_t2}}
\end{center}
\end{figure}

In this case the dynamics of the failure is driven by the most positive point. One might expect the most negative point to rapidly gain kinetic energy and overshoot the inflationary plateau, but this is only observed for higher values of $\Delta \phi$ and for small $\Delta\phi$ the field is ``pulled back'' by the gradients in the field. The most positive point in this case gets pulled back as before, but then hits the steep ``wall'' and proceeds to roll off the plateau. This is illustrated in Figure \ref{fig:SFphi_vs_t2}.

Again thinking solely about the scalar field dynamics of the extremal points, one can estimate at which point the most negative point will fail directly.  Consider the most negative value of $\phi$ initially, $\phi_{\rm min} = \phi_0 - 3 \Delta \phi$. We can see that it will fail if, having oscillated through $\phi_0$, the point at which it would be brought to rest by gradient pressure exceeds the critical point derived in the previous section of $\phi_{\rm max, crit} =0.00525\mpl$. If the potential were flat in this region, this value would be the same as $\phi_{\rm max} = \phi_0 + 3 \Delta \phi$ (since it is effectively in simple harmonic motion). However, the initial slope in $V(\phi)$ gives it an extra ``push", which we can equate to having started with a larger value of $\Delta \phi$. Considering the initial energy density $\rho_0$ at the minimum point, relative to the point $\phi_0$
\begin{eqnarray}
\rho_0 = (V(\phi_{\rm min}) -V(\phi_0)) + \frac{1}{2} (\nabla\phi)^2 ,
\end{eqnarray}
then by making the approximations $\phi_0 \ll 3 \Delta \phi$ and $V(\phi_0) \approx V_0$, this becomes
\begin{eqnarray}
\rho_0 = m^2 (-3 \Delta \phi)^2 + \frac{6 \pi^2 n^2 \Delta \phi^2 }{L^2} ,
\end{eqnarray}
and we can find an ``effective'' initial value of $\Delta \phi$, which would have the same initial energy density
\begin{eqnarray}
\Delta \phi_{\rm eff} = \Delta \phi \sqrt{1+ \frac{3 m^2 L^2}{2 \pi^2 n^2}} .
\end{eqnarray}
Setting this equal to $\phi_{{\rm max, crit}}$ gives a rough estimate for the initial value of $\Delta \phi$ leading to immediate failure at the minimum point, which using our specific values gives $\Delta \phi \approx 0.0002$. As shown in Figure \ref{fig:SFphi_vs_t2}, this value is consistent with our findings, although failure will occur slightly below this value, due to the various assumptions made, in particular that the potential is flat after $\phi_0$, when it is in fact sloped downwards. 

\section{Large Field Inflation} \label{sect:LF}

In large field inflation the inflationary part of the potential is $\delta \phi \gg \mpl$. It thus supports larger fluctuations in the field while still keeping the entire space within the inflationary regime.

Work to test the robustness of inflation in large field inflation was done in~\cite{KurkiSuonio:1993fg}, and more recently by \cite{East:2015ggf} who found that large field inflation with uniformly expanding initial conditions is very robust to large inhomogeneities of up to $\rho_{\rm grad} = 1000\rho_{V_0}$. In this section we broadly reproduce their results, before extending the work to consider the limit of very large fluctuations and non-uniform initial expansion rates which include initially contracting regions. 

We use $m^2 \phi^2$ inflation as a generic model for large scale inflation\footnote{While this model is marginally ruled out by the latest {\em Planck} data \cite{Ade:2015lrj}, we chose it for its simplicity of implementation. More complicated models will not lead to any drastically different results since the key feature is the flatness of the potential and the long traverse to the minimum.}
\begin{eqnarray}
V(\phi) = m^2 \phi^2 
\end{eqnarray}
with $m=1.07967 \times 10^{-7} \mpl$, leading to an inflationary scale of $H_{\rm inf} = 1.25 \times 10^{-6} \mpl$. For an initial value of the field of $\phi_0 = 4\mpl$, these values would result in 100 e-folds of inflation, with the scalar perturbation amplitude $\Delta_R = 10^{-5}$ and the scalar index $n_s \approx 0.97$ for modes that exit the horizon 60 $e$-folds before the end of inflation. 

The length scale for the fluctuations $L$ is set to the Hubble length in the absence of fluctuations, for our choice of parameters $L=8.0 \times 10^5 \mpl^{-1}$.

\subsection{Large field inflation with constant $K$}

In this section the value of $K$, the extrinsic curvature, is set as a constant across the grid using \eqn{eqn:Kavg} as above. We first considered a range of initial amplitudes of the perturbations $\Delta \phi$, and found that values below $0.2 \mpl$ resulted in inflation everywhere.  

We find that at larger values of $\Delta \phi$, we form black holes. As in \cite{East:2015ggf}, we can argue for their formation at this scale using the hoop conjecture. The black hole mass $M$ must be enclosed within a hoop of radius $R$. Assuming spherical symmetry, this is
\begin{eqnarray} 
R = 2GM .
\end{eqnarray}  
For a perturbation of wavelength $L$, given that a single mode in each spatial direction necessarily creates two  black holes by symmetry, the greatest radius from which each black hole can accrete is approximately
\begin{eqnarray} 
R = \frac{L}{4} \label{eqn:BHmass}.
\end{eqnarray}
Note that, $L$ is an arbitrary length (the wavelength of the perturbation) and not necessarily the Hubble length. The mass enclosed, $M$, is
\begin{eqnarray} 
M \approx \frac{4}{3} \pi R^3 \langle \rho_{\mathrm{grad}} \rangle
\end{eqnarray}
where
\begin{eqnarray} 
\langle \rho_{\mathrm{grad}}\rangle  \approx  3 \pi^2 \frac{\Delta \phi^2}{L^2}  \label{eqn:rhograd}
\end{eqnarray}
which is obtained from the volume average of $(\nabla \phi)^2$ from \eqn{eqn:phi} over a volume $L^3$. The \emph{maximum} mass is then linearly proportional to $L$, i.e.
\begin{equation}
M = \frac{\pi^3}{16} L \left(\frac{ \Delta \phi}{\mpl}\right)^2. \label{eqn:MassBH}
\end{equation}
Combining these gives the condition
\begin{eqnarray} 
\frac{\Delta \phi}{\mpl} \geq \sqrt{\frac{2}{\pi^3}}
\end{eqnarray}
as the critical case for black hole formation, independent of the length $L$. This value of approximately $0.25\mpl$ is consistent with our findings above that the critical $\Delta \phi \approx 0.2$ (given the approximate nature of the calculation).

Using equation~\eqn{eqn:equiv}, this value of $\Delta \phi \approx 0.25\mpl$ gives $k/H \approx 1.6$.  This result is also consistent with the findings of \cite{East:2015ggf}.  In other words, black holes will form when the wavelength of the perturbation is four times the Hubble length $H^{-1} \approx 3/\sqrt{24\pi \rho_{\mathrm{grad}}} \mpl$ or larger. 

\begin{figure}
\begin{center}
\includegraphics[width=8cm]{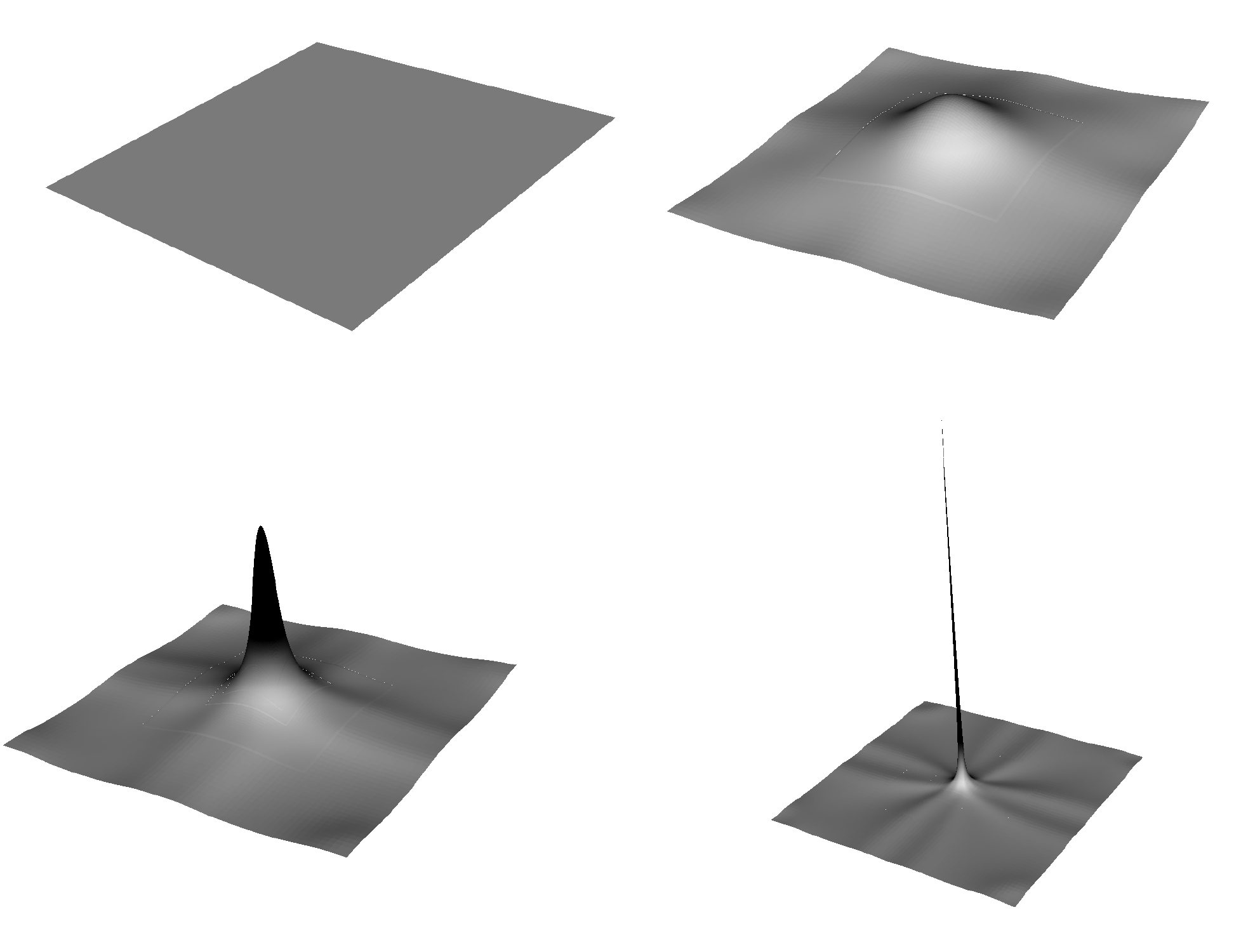}
\caption{A time series of the value of the extrinsic curvature $K$ is shown on a 2 dimensional spatial slice in the case $\Delta \phi = 0.5 \mpl$. Black areas are collapsing (within the black hole) and the remaining areas are inflating, with the colour scale from black to white varying between $K = \pm 5 \times 10^{-5} \mpl$. We confirm the formation of a black hole by using an apparent horizon finder, but are able to continue evolving the black hole until it ``inflates out" of the space. Although this is in part due to the gauge conditions (the moving puncture method tends to move the coordinate points away from the black hole singularity), the dominant effect is the inflation of the surrounding space. Eventually the spike will disappear once it falls below the coordinate grid resolution. 
\label{fig:LF_BH}}
\end{center}
\end{figure}

Above the critical value, black holes were formed, but these only created locally collapsing regions, and did not dominate the overall inflationary behaviour. As such they were quickly ``inflated out" of the spacetime, see Figure \ref{fig:LF_BH}.

The robustness was, as in the small field case, due in part to the fact that the most extreme value is quickly ``pulled back'' up the hill by gradient energy, resulting in initial inhomogeneities being smoothed out.  For the $m^2\phi^2$ potential we are using,  $\phi_0$ and $\Delta \phi$ are both large and so the field will move towards the minimum when
\begin{eqnarray}
\frac{32 \pi^3 n^2 m^2 \phi_0^2 \Delta \phi}{\mpl^2} < 2 m^2 (\phi_0 - 3 \Delta \phi),
\end{eqnarray}
which reduces to
\begin{eqnarray}
\Delta \phi < \frac{\phi_0\mpl^2}{16 \pi^3 n^2 \phi_0^2+3\mpl^2}\approx\frac{\mpl^2}{16\pi^3 n^2\phi_0}.
\end{eqnarray}
Thus there is a \emph{minimum} value beyond which pullback always occurs. This is because $dV/d\phi$ approaches zero at the minimum for $m^2\phi^2$ type potentials -- i.e. it is a convex potential. For concave type potentials (e.g. hill-top models \cite{Boubekeur:2005zm}), there will be a maximum $\Delta \phi$ instead, as we have discussed in the small field case. The value of this bound is small, in our model $\Delta \phi \approx 0.0005$, and in the limit of a very flat, extended potential, it is zero. Thus almost any perturbations will tend to be pulled back to a (potentially) more homogeneous configuration in this potential. This more homogeneous configuration then continues to inflate, with the number of e-folds approximately equal to that given by $\phi_0$. 

Thus, as expected, inflation eventually wins out, even with fluctuations which reach almost to the minimum of the $m^2 \phi^2$ potential -- \emph{large field inflation is very robust to scalar field inhomogeneities}. 

\subsection{Can Black Holes stop inflation?}

Naively, one might imagine that one can continuously increase the size of the fluctuations to generate black holes of increasing mass, to the point that the Schwarzschild radius dominates over the scale of inflation $R_{S} \gg H_V^{-1}$, ending inflation. The critical limit for this to occur is the so-called Nariai limit, which occurs when the black hole horizon and the de Sitter horizon coincide in a Schwarzschild-de Sitter spacetime. In our units, the critical mass of the black hole is
\begin{equation}
M_{\rm N} = \frac{1}{3}\frac{1}{\sqrt{8\pi}}\frac{\mpl^2}{\sqrt{V_0}}\mpl \label{eqn:nariai}.
\end{equation}
We will show in this section that this is not possible. 

\begin{figure}
\begin{center}
\includegraphics[width=8cm]{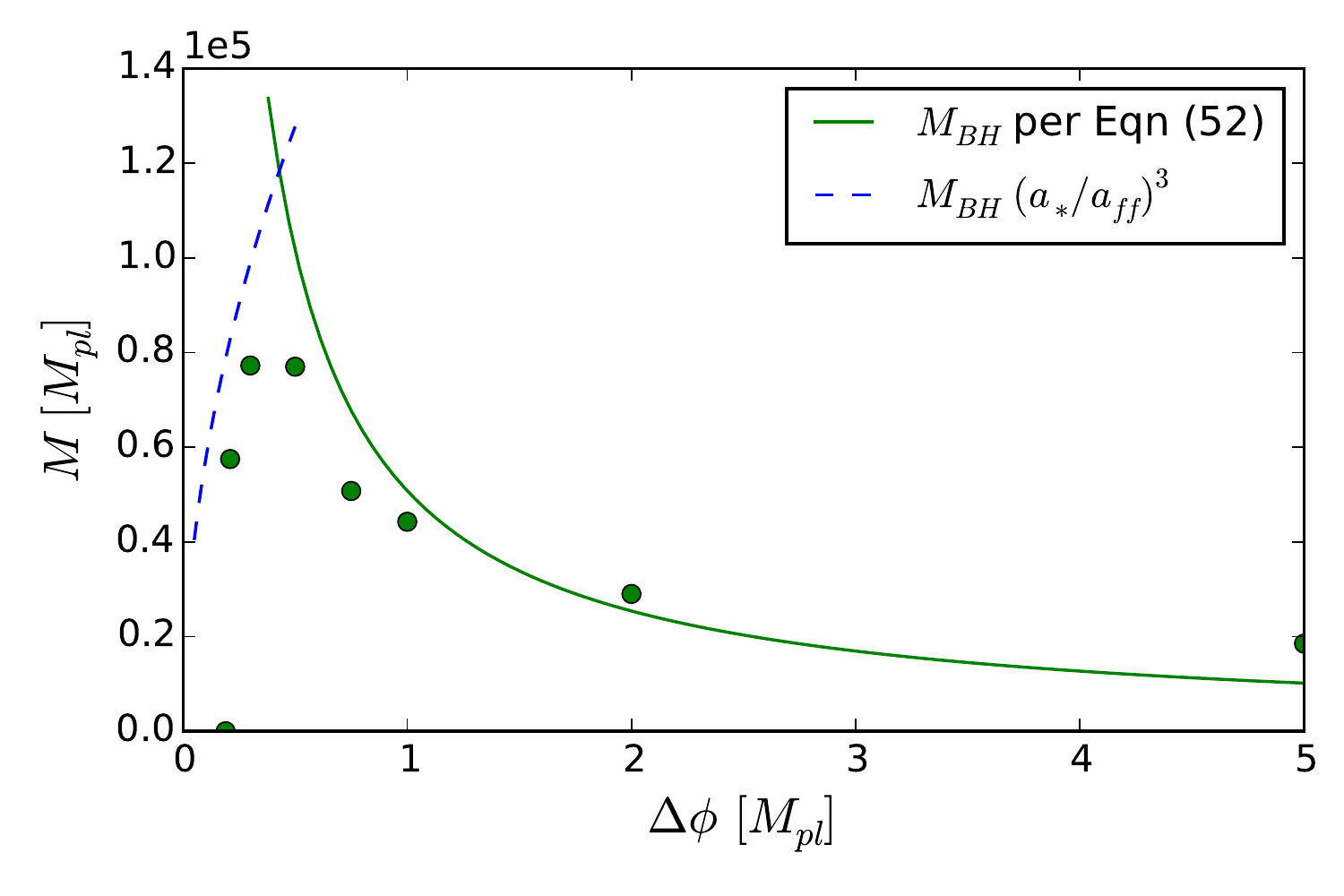}
\caption{Plot showing the mass of the black holes formed versus the size of the initial perturbations. Although the mass initially increases, it reaches a maximum (at a size of about $R_{\rm S}=0.2L$) after which it falls off as $1/\Delta\phi$, meaning the maximum mass of the black hole which can be formed is bounded. The green dots are results from numerical simulations, and the lines marked illustrate the approximate agreement to our analytic calculations. The green solid line shows the maximum mass $M_{\rm BH}$ predicted per \eqn{eqn:LFBHmass}. The blue dotted line shows $M_{\rm BH} (a_*/a_{\rm ff})^3$, reflecting the effect of the transition to de Sitter on the radius from which energy can fall in. The two lines meet at the critical point $\Delta \phi \approx 0.43 \mpl$.
\label{fig:LFBHMass}}
\end{center}
\end{figure}

In Figure \ref{fig:LFBHMass}, we show the mass of the black holes formed as we gradually increase the amplitude $\Delta \phi$  obtained from numerical simulations. We see that while the mass increases with $\Delta \phi$ initially, at some point, the black hole mass begins to \emph{decrease} as $M \propto \Delta \phi^{-1}$. Thus there is a maximum black hole mass which can be formed (which in our case had a Schwarzschild radius of about 20 per cent of the Hubble radius related to the initial $V(\phi)$, that is, $L$).  This can be understood as follows.

Since $\rho_{\mathrm{grad}} \gg V_0$ initially, the early expansion of the spacetime is roughly that of a radiation dominated universe, i.e. $\rho_{\mathrm{grad}} \propto a^{-4}$ and $H^2 \propto \rho_{\mathrm{grad}}$.  At late times, the expansion rate is that of de Sitter, i.e. $a\propto e^{H_V t}$ where $H_V^2 = (8\pi G/3)V_0$. 

Meanwhile, the free fall time-scale for some matter distribution of average density $\rho$ is given by
\begin{equation}
\Delta t_{\rm ff} \approx \sqrt{\frac{1}{G \rho}} \label{eqn:tff}.
\end{equation}
If there is no expansion, then this is roughly the timescale for some cloud of density $\rho$ to collapse to form a black hole as long as the initial distribution is supercritical. However, due to the presence of the large gradient energy density, the spacetime is roughly expanding as a radiation dominated universe, dissipating some of the energy away from forming a black hole. Once the spacetime is dominated by vacuum energy, it is safe to assume that any remaining energy that has not collapsed into a black hole will be rapidly dissipated. The time scale for this to happen, $a_*$, occurs at vacuum-gradient energy equality $\rho_{\mathrm{grad}}a_*^{-4} = V_0$, i.e.\footnote{Recall that in our conventions $a\approx 1$ on the initial slice.}
\begin{equation}
a^2_* = \sqrt{8\pi^3} \frac{\Delta \phi}{\mpl} \label{eqn:astar}\,,
\end{equation}
where we have used \eqn{eqn:Lformula} and \eqn{eqn:rhograd}. 

Converting $\Delta t_{\rm ff}$ into the scale factor by solving the Friedmann equation, we get
\begin{equation}
a_{\rm ff}^2 = 2\sqrt{\frac{8\pi}{3}} +1 \label{eqn:aff}\,,
\end{equation}
which is independent of $\rho_{\mathrm{grad}}$ as expected. This predicted value of $a_{\rm ff}=2.6$ provides a good approximation to the value of $\langle a \rangle \approx 3$ measured at black hole formation in the simulations over the range of $\Delta \phi$ tested. If $a_* < a_{\rm ff}$, then de Sitter space will take over before the collapse has finished, leading to a lower mass black hole, and this is the case for \emph{smaller} values of $\Delta \phi$.  By equating these two times from \eqn{eqn:astar} and \eqn{eqn:aff}, we obtain 
\begin{equation}
\Delta \phi \approx 0.43 \mpl\,,
\end{equation}
as the point when the free fall is no longer stopped by de Sitter expansion, resulting in a maximum mass of the black hole. This is in good agreement with the numerical value which gives the maximum mass at $\Delta \phi \approx 0.4 \mpl$, as seen in Figure \ref{fig:LFBHMass}.

Above this point, the free fall timescale falls fully within the radiation dominated era. Consider the mass enclosed in a spherical distribution of matter of size $r$
\begin{equation}
M(r) = \frac{4}{3}\pi r^3 \rho\,.
\end{equation}
Since the collapse occurs well within the radiation domination era, the largest radius from which matter can still collapse into a black hole is the Hubble radius, $r \sim H^{-1}$, with the largest mass occurring when $\rho = \rho_{\mathrm{grad}}$, giving us 
\begin{equation}
M_{\rm BH} \lsim \frac{4 \sqrt{3}}{(8\pi)^2} \left(\frac{\mpl}{\Delta \phi}\right) \left(\frac{\mpl^2}{\sqrt{V_0}}\right) \mpl, \label{eqn:LFBHmass}
\end{equation}
which scales like $M \propto \Delta \phi^{-1}$ as our numerical results indicate, and has a Schwarzschild radius of
\begin{equation}
R_{\rm S} \lsim \frac{L}{\sqrt{8\pi^3}} \frac{\mpl}{\Delta \phi},
\end{equation}
which agrees to the maximum observed size of about $R_{\rm S}=0.2L$ for $\Delta\phi=0.4$. This means that one cannot make a ``Giant Death Black Hole'' using the methods we outlined in this work -- there is a maximum mass, roughly $1/3$ of the mass of the Nariai black hole \eqn{eqn:nariai}, after which increasing $\Delta \phi$ leads to a reduction in BH size. While our analysis and numerical simulations have focused on the specific case where the initial expansion is uniform and scaled to the gradient energy, we expect similar no-go results to hold as long as the initial hypersurface is approximately flat  i.e. $\chi \sim 1$ and $A_{ij} \approx 0$ since the Hamiltonian constraint \eqn{eqn:HamCon} implies that the initial expansion will be, on average, uniform and large. It would be interesting to revisit this question in the more general case where these assumptions are relaxed.

\subsection{Large field inflation with spatially varying $K$}

We now consider spatially varying $K$ in the case where \mbox{$\Delta \phi=0.1\mpl$} and study the effect on inflation. The potential is now set to be simply a cosmological constant with the value $V(\phi_0)$ from the previous large field case, to allow inflation to continue indefinitely.

\begin{figure}
\begin{center}
\includegraphics[width=8cm]{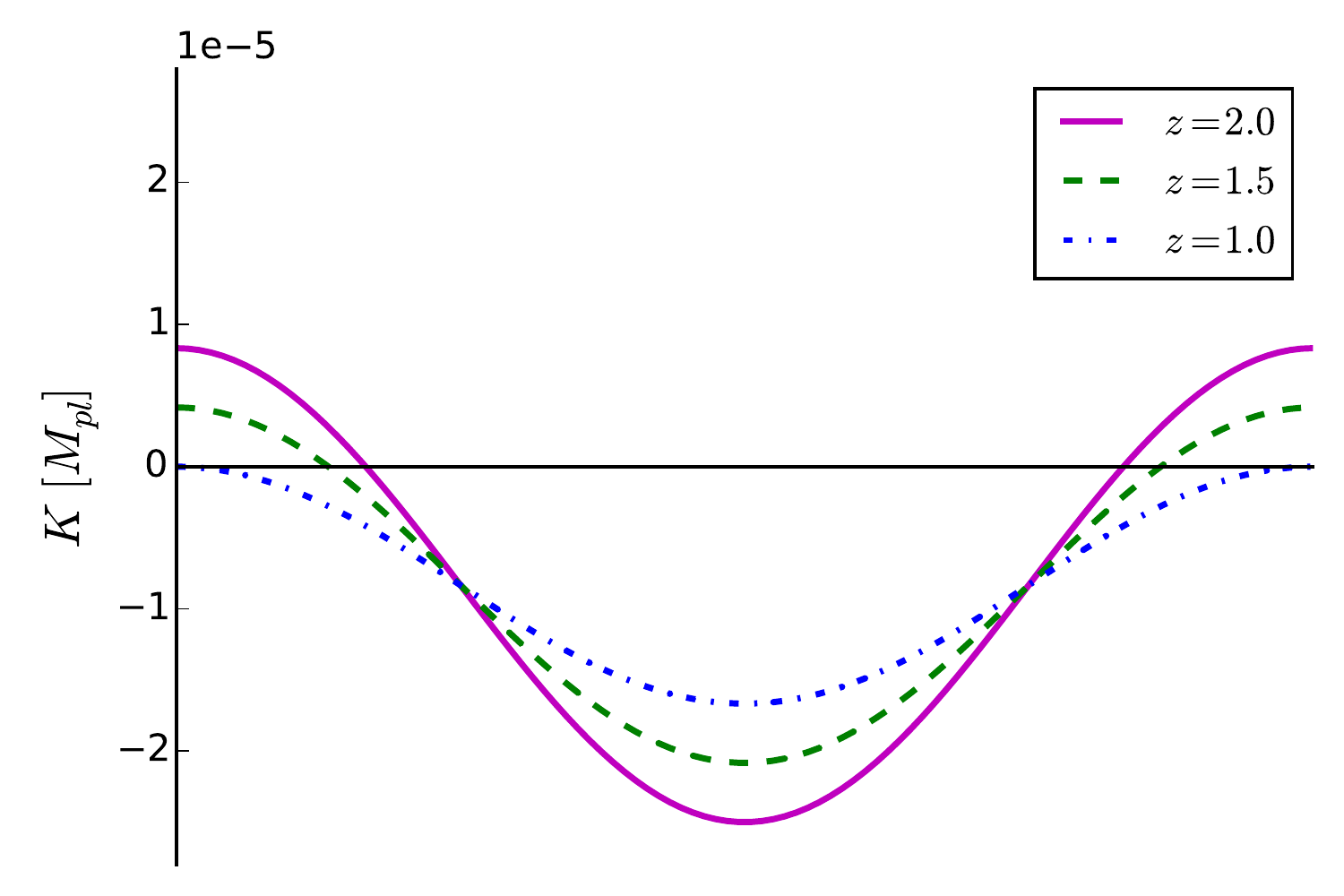}
\caption{Illustration of the change in the spatial variation in $K$ when the parameter $z$ is varied, showing a slice through the maximum and minimum values of the profile. 
\label{fig:Kprofiles}}
\end{center}
\end{figure}

For our purposes it is useful to recast \eqn{eqn:Kansatz} for $K$ in the form
\begin{equation}
K = - z \bar{C}(\phi-\phi_0) + \langle K \rangle \label{eqn:Kansatz2}\,.
\end{equation}
We set
\begin{equation}
\langle K \rangle  = -\sqrt{24 \pi G \langle \rho \rangle}\,, \label{eqn:Kavg2}
\end{equation}
where the value of $\rho$ now includes the contribution from $\eta$.
Without loss of generality, we set $\bar{C}=2.78 \times 10^{-5}$ so that the maximum value of $K$ is zero for $z=1$. Increasing $z$ increases the amplitude of the fluctuations in $K$ and allows us to consider larger regions of spacetime that are initially collapsing, $K>0$. The profile for $K$ and the dependence on $z$ is illustrated in Figure \ref{fig:Kprofiles}. 

We test a range of values of $z$ between $1.0$ and $2.0$, and find that in cases of smaller $z$ (where most of the spacetime is expanding initially) the collapsing part of the spacetime ``bounces back", such that $K$ quickly becomes approximately constant with a negative value everywhere. Inflation then continues, and over 20 efolds are reached. 

\begin{figure}
\begin{center}
\includegraphics[width=8cm]{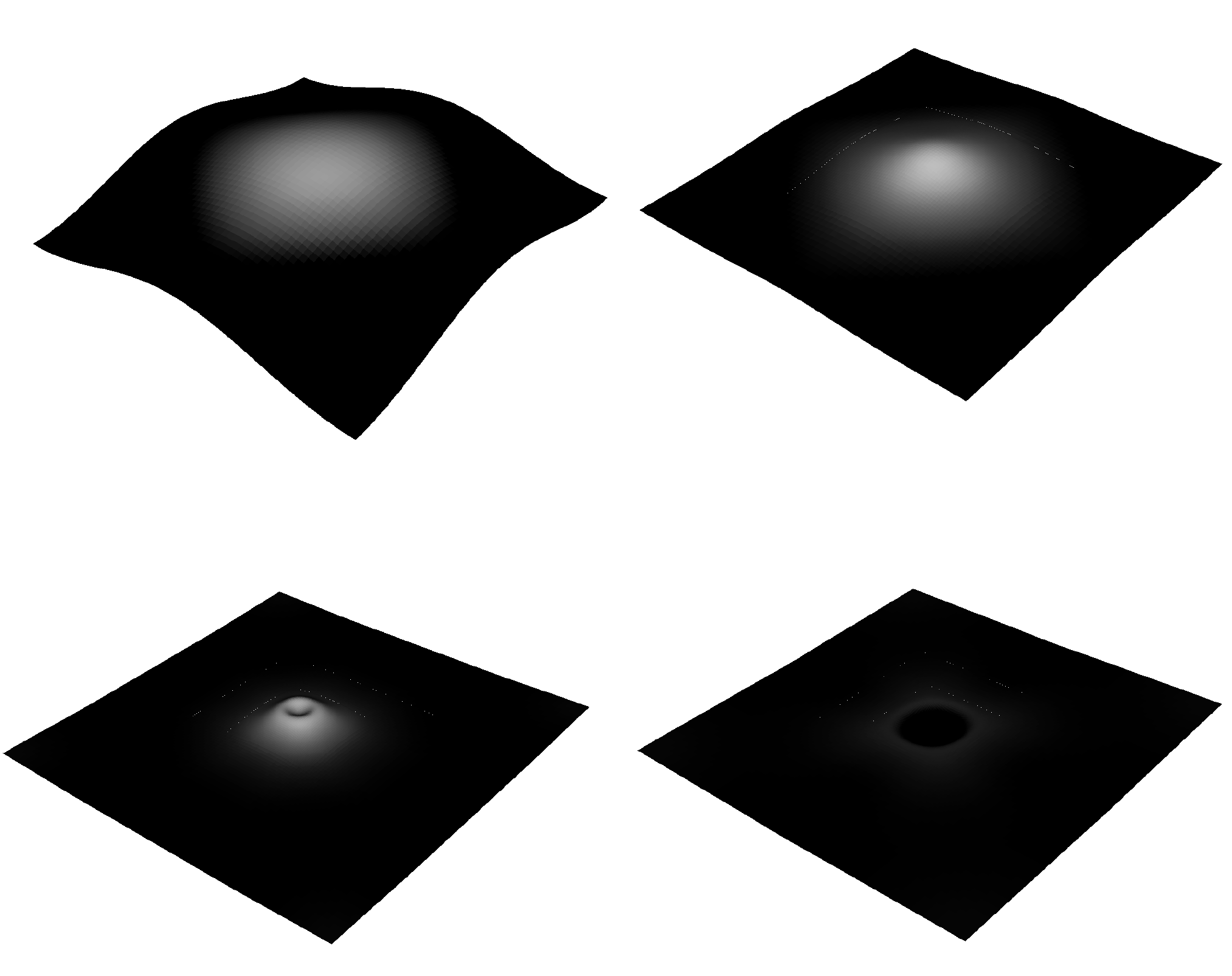}
\caption{Illustration of the evolution of $K$ in the large field varying $K$ case for $z=1.5$. The initial inhomogeneities in $K$ quickly disperse and it settles into an inflating spacetime everywhere. White areas are collapsing and black areas are inflating, with the colour scale from black to white varying between $K = \pm 5 \times 10^{-5} \mpl$. 
\label{fig:LFK_vs_xt}}
\end{center}
\end{figure}

\begin{figure}
\begin{center}
\includegraphics[width=8cm]{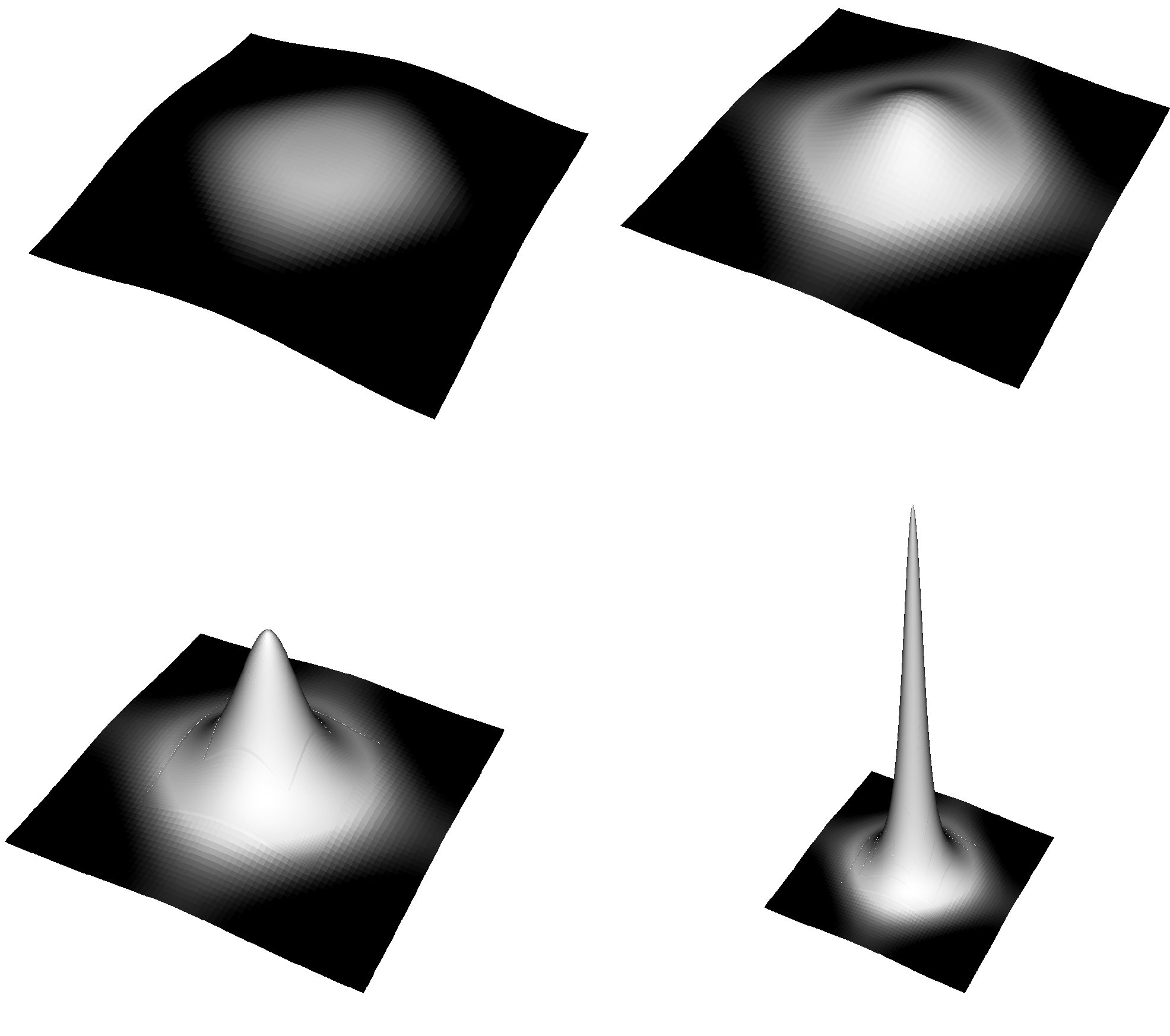}
\caption{Illustration of the evolution of $K$ in the large field varying $K$ case for $z=2.0$, showing the black hole formation at the centre. White areas are collapsing and black areas are inflating, with the colour scale from black to white varying between $K = \pm 5 \times 10^{-5} \mpl$. The peak value in the last frame is $K = 0.005 \mpl$, and the average number of e-folds across the grid at this point is roughly $0.4$. 
\label{fig:LFK_vs_xt2}}
\end{center}
\end{figure}

In cases of higher $z$, with $z > 2.0$, where more of the spacetime (but still less than half) is collapsing initially, we find that black holes form at the initially contracting point. We therefore find that for an inhomogeneous spacetime which would have resulted in inflation everywhere for constant $K$ ($\Delta \phi=0.1$ is subcritical for black hole formation in the constant $K$ case), we are now able to generate regions of collapse to form black holes once variations of $K$ are introduced.  This is illustrated in Figures \ref{fig:LFK_vs_xt} and \ref{fig:LFK_vs_xt2}.

However, even in these cases, the remaining spacetime continues to expand and inflate. Since $\langle K \rangle <0$ in all cases here, this result is consistent with what would be expected from \cite{Barrow:1985} and \cite{Kleban:2016sqm} (note that our sign convention means that $K<0$ denotes locally expanding spacetime). We will explore the case where $\langle K \rangle >0$ in future work.

\section{Conclusions} \label{sect:conclusions}

We investigated the robustness of small and large field models of inflation, subjecting it to several simple inhomogeneous initial conditions both in the scalar field profile and in the extrinsic curvature. In doing so we have set up a framework that will allow us to study more general initial conditions in the future. As expected, we found that large field inflation was far more robust than small field inflation. In particular, small field inflation can fail even for small subdominant gradient energies $\rho_{\mathrm{grad}}/\rho_{V} \approx 10^{-4}$ while large field inflation is robust even to dominant gradient energies of $\rho_{\mathrm{grad}}/\rho_{V}  \gg 1$. This implies that small field inflation requires at least some level of tuning to begin or a dynamical mechanism that sets up appropriate initial conditions.

\subsection{Robustness of Small Field Inflation}

The primary failure mode for small field inflation is the disruption of coherent slow roll dynamics, causing some parts of the scalar field to irrevocably fall into the non-inflating minimum. Once a region of the scalar field falls into the minimum, this region will expand and dominate the rest of spacetime ending inflation for the entire hypersurface. This failure mode can be induced in the following ways:

\begin{itemize}
\item \emph{Adding large amplitude scalar fluctuations.} Large amplitude scalar fluctuations can create excursions outside the inflationary part of the potential which lead to one region falling to the minimum and dragging the rest of the field down with it. However, local excursions of the field towards the edge of the inflationary part of the potential get pulled back by gradient pressure, making small field inflation more robust than one might expect (see below).
\item \emph{Converting additional potential energy into kinetic energy.} If the inflationary region of the potential is small -- in the case of typical small field models it is often just an inflection point -- then a large initial fluctuation may have support on the steep part of the potential (i.e. the green dotted line in Figure \ref{fig:PotSFsteep}). This additional potential energy will be converted to scalar kinetic energy, generating a large fluctuation and pushing the scalar closer to the minimum, thus ending inflation.
\end{itemize}

Nevertheless, we found that small field inflation is more robust than one might naively expect. In particular, we find the following:

\begin{itemize}
\item \emph{Pullback effect of gradients.} We show that perturbations tend to ``homogenise'', i.e. gradients tend to flatten out. This means that some initial conditions which have regions in the non-inflating regime can still inflate as the scalar field gets pulled back into the inflating regime. We provide a formula for this critical point for any given potential, and demonstrate its effect numerically.
\item \emph{Adding additional shorter wavelength modes makes inflation more robust for a given maximum initial value of $\phi$.} Adding a second mode with half the wavelength, but normalised to keep the same value of $\phi_{\rm max}$, resulted in a higher threshold for inflation. This is somewhat unexpected since adding an $N=2$ mode \emph{increases} the gradient energy. We proposed that this could be related to the pullback effect described above, which is stronger for higher wavenumber modes.
\end{itemize}

\subsection{Robustness of Large Field Inflation}
In the large field case, except for the trivial case of an initial hypersurface which is contracting everywhere, i.e. $K > 0$, we did not find a viable failure mode for the initial conditions we considered -- large field inflation is robust to very large gradient energies $\rho_{\mathrm{grad}}/\rho_V > 10^2$. The primary reason for its robustness is the potential's large support for slow roll i.e. $\delta \phi \gg \mpl$, which combined with the rapid dissipation of gradient energy due to expansion, makes it difficult for the scalar to reach a non-inflating region. Furthermore, we find the following:

\begin{itemize}
\item \emph{No ``Giant Death Black Holes''}. Given a uniformly expanding initial condition scaled to the total initial gradient energy, there is a maximum mass black hole that is formed for which the radius is of order $0.2$ times the size of the vacuum energy Hubble radius. Increasing initial gradients beyond this point \emph{decreases} the final black hole mass -- this is caused by the dissipation of gradient energies due to the large initial expansion. We calculate the maximum mass, which occurs when the transition to de Sitter expansion no longer limits the black hole mass, and confirm it with numerical simulations. We find that it is roughly $1/3$ of the mass of the Nariai black hole.
\item \emph{Pullback effect of gradients}. Similar to the small field model, large gradients tend to homogenise. We show that for convex potentials, even with initial fluctuations which reach to the minimum of the potential, inflation can eventually succeed.
\item \emph{If $\langle K \rangle <0$, then inflation wins}.  If the initial hypersurface (a Cauchy surface) has a net negative (expanding) value of $K$ there will always be an expanding region, as predicted in analytic studies \cite{Barrow:1985} and \cite{Kleban:2016sqm}.
\end{itemize}

As we stated in the Introduction, we only investigated a very restrictive class of initial conditions.  Furthermore, we focused on the case of single field inflation. It will be interesting to study the effects of more general initial conditions and whether the presence of additional degrees of freedom renders inflation more or less robust to inhomogeneities.  We will pursue these and other questions in future work.

\acknowledgments

We thank Jonathan Braden, William East, Richard Easther, Matt Kleban, Hiranya Peiris, Matt Johnson, and Richard Matzner for useful conversations, and Robert Brandenberger for comments on our first draft. We would also like to thank the GRChombo team (http://grchombo.github.io/collaborators.html) for their work on the code, and various useful insights, and the COSMOS team at DAMTP, Cambridge for their ongoing technical support. Numerical simulations were performed on the COSMOS supercomputer, part of the DiRAC HPC, a facility which is funded by STFC and BIS. EAL acknowledges support from an STFC AGP grant ST/L000717/1. This work also used the ARCHER UK National Supercomputing Service (http://www.archer.ac.uk) for some simulations. BSD is supported in part by the National Science Foundation under grant PHY-1521186 and Maxwell Analytics LLC. BSD thanks the Helsinki Institute of Physics and the University of Amsterdam Institute for Theoretical Physics for their hospitality during part of this project. WF is supported by the National Science Foundation under Grant Number PHY-1316033. RF is supported in part by the Alfred P. Sloan Foundation. SP is grateful for support provided by the National Science Foundation under Grant Number PHY-1521186. 


\bibliography{mybib.bib}

\end{document}